\documentclass[onecolumn,preprintnumbers,amssymb,preprint,nofootinbib]{revtex4}
\usepackage[utf8]{inputenc} 
\usepackage[english]{babel}
\usepackage[T1]{fontenc}
\usepackage{graphicx,amssymb,amsmath,wasysym}
\usepackage[bookmarks=false]{hyperref}
\usepackage{graphicx}
\usepackage{dcolumn}
\usepackage{bm}

\def\gtap{\mathrel{ \rlap{\raise 0.511ex \hbox{$>$}}{\lower 0.511ex
   \hbox{$\sim$}}}} 
\def\ltap{\mathrel{ \rlap{\raise 0.511ex
    \hbox{$<$}}{\lower 0.511ex \hbox{$\sim$}}}} 
\newcommand{\bea}{\begin{eqnarray}} 
\newcommand{\eea}{\end{eqnarray}}
\def\beq{\begin{equation}}
\def\enq{\end{equation}}
\def\ba{\begin{eqnarray}}
\def\ea{\end{eqnarray}}
\def\<{<\!\!}
\def\>{\!\!>}
\def\<{\langle}
\def\>{\rangle}

\begin{document}

\input{epsf}

\begin{flushright}
Bonn-TH-2012-19\\
IFIC/12-57 \\
CFTP/12-012 \\
\end{flushright}
                 
\title{A novel way of constraining WIMPs annihilations in the Sun:
  \\ MeV neutrinos \vspace{5mm}}

\author{\large Nicol\'as Bernal$^1$, Justo Mart\'{\i}n-Albo$^2$ and
  Sergio Palomares-Ruiz$^3$} 

\affiliation{\vspace{5mm}$^1$Bethe Center for Theoretical Physics and
  Physikalisches Institut, Universit\"at Bonn, Nu\ss allee 12, D-53115
  Bonn, Germany \vspace{2mm}}
\affiliation{$^2$Instituto de Física Corpuscular (IFIC), 
CSIC-Universitat de València, Apartado de Correos 22085, E-46071
Valencia, Spain \vspace{2mm}}
\affiliation{$^3$Centro de Física Teórica de Partículas (CFTP),
Instituto Superior Técnico, Universidade Técnica de Lisboa, Av.
Rovisco Pais 1, 1049-001 Lisboa, Portugal \vspace{1cm}}

\begin{abstract}
Annihilation of dark matter particles accumulated in the Sun would
produce a flux of high-energy neutrinos whose prospects of detection
in neutrino telescopes and detectors have been extensively discussed
in the literature.  However, for annihilations into Standard Model
particles, there would also be a flux of neutrinos in the MeV range
from the decays at rest of muons and positively charged pions.  These
low-energy neutrinos have never been considered before and they open
the possibility to also constrain dark matter annihilation in the Sun
into $e^+e^-$, $\mu^+\mu^-$ or light quarks.  Here we perform a
detailed analysis using the recent Super-Kamiokande data in the few
tens of MeV range to set limits on the WIMP-nucleon scattering cross 
section for different annihilation channels and computing the
evaporation rate of WIMPs from the Sun for all values of the
scattering cross section in a consistent way.
\end{abstract}

\maketitle

\section{Introduction}
\label{sec:intro}
 
There is overwhelming evidence of the existence of a massive
non-baryonic dark component which contributes to about 80\% of the
energy budget of the Universe~\cite{Jungman:1995df, Bergstrom:2000pn, 
  Munoz:2003gx, Bertone:2004pz, Bertone:2010}, being a weakly
interacting massive particle (WIMP), with mass lying from the GeV to
the TeV scale, one of the most popular candidates.

One of the different proposed strategies to detect WIMPs is to search
for the flux of high-energy neutrinos from the annihilations of WIMPs
accumulated in the center of the Sun~\cite{Silk:1985ax,
  Srednicki:1986vj, Hagelin:1986gv, Gaisser:1986ha}.  Many different
studies have evaluated the prospects of detection of these neutrinos
with neutrino telescopes/detectors~\cite{Kamionkowski:1991nj,
  Bottino:1991dy, Halzen:1991kh, Bergstrom:1996kp, Barger:2001ur,
  Cirelli:2005gh, Mena:2007ty, Blennow:2007tw, Wikstrom:2009kw,
  Rott:2011fh, Das:2011yr}.  However, previous works have focused on
WIMPs annihilations into hadronic or $\tau^+\tau^-$ channels.  On the
other hand, annihilations into $\mu^+\mu^-$ or light quarks have
always been neglected, for muons and pions lose energy very
effectively in the dense regions where they would be produced and then
would decay at rest, giving rise to neutrinos in the MeV range.
Likewise, annihilations into $e^+e^-$ have never been considered, for
they would not produce directly neutrinos.  Nevertheless, in their
propagation through the Sun they would interact with nuclei and
produce pions, which would be stopped.  The $\pi^-$ would then get
captured and subsequently absorbed by the nuclei of the medium, but
the $\pi^+$ would decay at rest, producing a flux of MeV neutrinos.
On the other hand, hadronic and $\tau^+\tau^-$ channels, along with
heavy mesons (the source of the high-energy neutrinos considered so
far), would also produce light mesons, as pions, which would then be
stopped and (in the case of $\pi^+$) decay at rest.  The energies of
these neutrinos lie at the energy range where the diffuse supernova
neutrino background (DSNB) is searched for by detectors such as
Super-Kamiokande (SK)~\cite{Bays:2011si, Baysthesis}\footnote{It is
  interesting to note that this is also the energy region for GUT
  monopole searches at SK~\cite{Ueno:2012md}, which have a spectral
  signal of the same type of the one discussed in this work.  However,
  in that analysis only angular bins were considered, whereas in this
  work we make use of the full energy spectrum.}.

Here we consider, for the first time\footnote{This idea was
  simultaneously proposed by Ref.~\cite{Rott:2012qb}.  Both works were
  made publicly available on the arXiv the very same day.}, the
potential signal of these low-energy neutrinos from WIMPs
annihilations in the Sun and use the most recent SK
data~\cite{Bays:2011si, Baysthesis}, and analogously to the SK
collaboration, we perform an extended maximum likelihood analysis in
order to set bounds on the scattering cross section of WIMPs off
nucleons for different annihilation channels.  In this work we
calculate the evaporation rate of WIMPs from the Sun for all values of
the scattering cross section in a consistent way and note that in the
optically thick regime, it {\it decreases} with the cross section,
which allows us to set limits to WIMP masses usually not considered
within this context.

\section{Capture, annihilation and evaporation of WIMPs in the Sun}
\label{sec:basics}

Galactic WIMPs would get eventually trapped in the Sun if, after
many elastic scatterings off the solar nuclei, they lose energy and
their velocity gets lower than the Sun's escape velocity.  If the
mean free path of WIMPs is large compared to the size of the Sun (the
Knudsen limit or optically thin regime), they would thermalize
non-locally by multiple interactions, so their density could be
approximated as an isothermal sphere following the law of atmospheres,
with a radial dependence set by the gravitational
potential~\cite{Spergel:1984re, Faulkner:1985rm}
\begin{equation}
\label{Eq:WIMPdistiso}
  n_{\chi(r), {\rm iso}} (r, t) = N_{\chi}(t) \, \frac{e^{-m_\chi
    \phi(r)/T_\chi}}{\int_{0}^{R_\odot} 4\pi r^2 dr \, e^{-m_\chi
      \phi(r)/T_\chi}} ~,
\end{equation}
where $N_{\chi} (t)$ is the total population of WIMPs with mass $m_\chi$,
$\phi(r)=\int_0^r G M_\odot(r')/r'^2 dr'$ the solar gravitational
potential at $r$ and $T_\chi$ the average WIMPs temperature,
calculated by imposing that there is no net flow of 
energy~\cite{Spergel:1984re} and using the Standard Solar Model
(SSM)~\cite{Asplund:2009fu, Serenelli:2011py} (we consider 29
elements).

However, for large cross sections (optically thick regime), WIMPs
would be in local thermal equilibrium and their density distribution
could be approximated as~\cite{Nauenberg:1986em, Gould:1989hm}
\begin{equation}
\label{Eq:WIMPdistLTE}
  n_{\chi(r), {\rm LTE}} (r, t) = n_{\chi, {\rm LTE}}(0,t) \,
  \left(\frac{T_\odot(r)}{T_\odot(0)}\right)^{3/2} \, {\rm
    exp}\left(-\int_{0}^{r} \frac{\alpha(r')
    \frac{dT_\odot(r',t)}{dr'} + m_\chi
    \frac{d\phi(r')}{dr'}}{T_\odot(r')} dr'\right) ~,
\end{equation}
where $T_\odot(r)$ is the solar temperature at radius $r$ and $n_{\chi,
  {\rm LTE}}(0,t)$ is set by the normalization $\int_{0}^{R_\odot}
4\pi r^2 dr \, n_{\chi, {\rm LTE}} (r, t) = N_\chi (t)$.  The factor
$\alpha (r)$ is the dimensionless thermal diffusivity and, for a given
admixture of elements in the medium, a good approximation is to take
the weighted mean of the solutions to the single-element
case~\cite{Gould:1989hm, 1990ApJ...356..302G},
\begin{equation}
\label{Eq:alpha}
\alpha(r) = \ell(r) \sum_i \ell_i(r)^{-1} \, \alpha_0 (m_i/m_\chi) ~,
\end{equation}
where $\alpha_0$ is the diffusivity for one element and is tabulated
as a function of $m_i/m_\chi$ in Ref.~\cite{Gould:1989hm}, where $m_i$
is the mass of the $i$-th nuclear species.  The quantity $\ell(r)=
(\sum_i \ell_i(r)^{-1})^{-1}$ is the total mean free path of WIMPs and
$\ell_i(r) = (\sigma_i n_i(r))^{-1}$ is the partial mean free path for
WIMP interactions with cross section $\sigma_i$ off the $i$-th nuclear
species with density $n_i(r)$, for which we use the
SSM~\cite{Asplund:2009fu, Serenelli:2011py}.  Depending on the type of  
interactions, either spin-dependent or spin-independent, the cross
section is given by
\begin{eqnarray}
\label{Eq:csSD}
\sigma_i^{\rm SD} & = & \left(\frac{\mu_i}{\mu_{\rm p}}\right)^2 \,
\frac{4 (J_i +1)}{3 J_i} \, \left| \langle S_{{\rm p}, i}\rangle +
     {\rm sign}(a_{\rm p} a_{\rm n}) \left(\frac{\mu_p}{\mu_n}\right) \, 
\sqrt{\frac{\sigma_{\rm n}^{\rm SD}}{\sigma_{\rm p}^{\rm SD}}} \langle
S_{{\rm n},i} \rangle \right|^2 \, \sigma_{\rm p}^{\rm SD} ~, \\  
\label{Eq:csSI}
\sigma_i^{\rm SI} & = & \left(\frac{\mu_i}{\mu_p}\right)^2 \,
 \left| Z_i + (A_i-Z_i) \, {\rm sign}(f_{\rm p} f_{\rm n}) \,
\left(\frac{\mu_p}{\mu_n}\right) \, \sqrt{\frac{\sigma_{\rm n}^{\rm
      SI}}{\sigma_{\rm p}^{\rm SI}}} \right|^2 \, \sigma_{\rm p}^{\rm
  SI} ~,
\end{eqnarray}
where $\mu_i$ ($\mu_{p/n}$) is the reduced mass of the WIMP-nucleus $i$
(WIMP-proton/neutron) system, $\sigma_{\rm p}^{\rm SD}$ ($\sigma_{\rm
  n}^{\rm SD}$) and $\sigma_{\rm p}^{\rm SI}$ ($\sigma_{\rm n}^{\rm
  SI}$) are the spin-dependent and spin-independent elastic scattering
WIMP cross section off protons (neutrons), respectively, $Z_i$, $A_i$
and $J_i$ are the atomic number, the mass number and the spin of the
nucleus $i$, and $\langle S_{{\rm p},i}\rangle$ and $\langle S_{{\rm n},
  i}\rangle$ are the expectation values of the spins of protons and
neutrons averaged over all nucleons, respectively, which we take from
Refs.~\cite{Ellis:1987sh, Engel:1989ix, Engel:1992bf, Divari:2000dc}.
The quantities $a_{\rm p}$ ($f_{\rm p}$) and $a_{\rm n}$ ($f_{\rm n}$)
are the axial (scalar) four-fermion WIMP-nucleon couplings.  As usual,
we assume $\sigma_{\rm p}^{\rm SD} = \sigma_{\rm n}^{\rm SD}$,
$\sigma_{\rm p}^{\rm SI} = \sigma_{\rm n}^{\rm SI} \equiv \sigma^{\rm
  SI}$ and the same sign for the couplings, so Eqs.~(\ref{Eq:csSD}) 
and~(\ref{Eq:csSI}) get simplified as 
\begin{eqnarray}
\label{Eq:crosssectionsSD}
\sigma_i^{\rm SD} & = & \left(\frac{\mu_i}{\mu_{\rm p}}\right)^2 \,
\frac{4 (J_i +1)}{3 J_i} \, \left| \langle S_{{\rm p}, i} \rangle +
\langle S_{{\rm n},i} \rangle \right|^2 \, \sigma_{\rm p}^{\rm SD} ~,
\\  
\label{Eq:crosssectionsSD}
\sigma_i^{\rm SI} & = & \left(\frac{\mu_i}{\mu_{\rm p}}\right)^2 \,
 A_i^2 \, \sigma^{\rm SI} ~.
\end{eqnarray}
Nevertheless, in the case of spin-dependent cross section it is the
coupling with protons which is mainly probed, for almost all WIMPs
interactions are off hydrogen. 

The transition from one regime to the other is indicated by the
so-called Knudsen number, 
\begin{equation}
\label{Eq:Knudsen}
Kn \equiv \frac{\ell(0)}{r_\chi} ~,
\end{equation}
where $r_\chi = \left(\frac{3 \, T(0)}{2 \pi G \, \rho(0) \,
  m_\chi}\right)^{1/2} $ is the approximate scale height of the WIMP
distribution, with $G$ the gravitational constant and $\rho(0)$ the
density at the solar center.  In order to interpolate between the
optically thin ($Kn \gg 1$) and the optically thick ($Kn \ll 1$) regimes
we follow the approach of Ref.~\cite{Scott:2008ns}, motivated by the
results of Ref.~\cite{Gould:1989hm}, and approximate the total WIMP
distribution as 
\begin{eqnarray}
\label{Eq:WIMPdist}
n_{\chi} (r, t) & = & \mathfrak{f}(Kn) \, n_{\chi, {\rm LTE}} (r, t) +
\left(1-\mathfrak{f}(Kn)\right) \, n_{\chi, {\rm iso}} (r, t) \\[1ex]
&& \mathfrak{f}(Kn) =  1 - \frac{1}{1+(0.4/Kn)^2}  ~. \nonumber
\end{eqnarray}

The evolution of the total number of WIMPs in the Sun is governed by the
following equation: 
\begin{equation}
\label{Eq:evolution}
  \dot{N_\chi} (t) = C_\odot - A_\odot \, N_\chi^2(t) - E_\odot \,
  N_\chi (t) ~,
\end{equation}
where $C_\odot$ is the capture rate, $A_\odot$ is the annihilation
rate and $E_\odot$ is the evaporation rate, which is only relevant for
low-mass WIMPs.  

For weak cross sections, the capture rate is defined
as~\cite{Gould:1987ju, Gould:1987ir} 
\begin{equation}
\label{Eq:captureweak}
C_\odot^{\rm weak} = \sum_i \int_0^{R_\odot} 4\pi r^2 dr \,
\int_0^{\infty} du \, \left(\frac{\rho_\chi}{m_\chi}\right) \,
\frac{f_{v_\odot}(u)}{u} \, \omega(r) \, \int_0^{v_e} R_i^- (\omega
\rightarrow v) |F_i(\omega,v)|^2\, dv  ~, 
\end{equation}
where $R_i^- (\omega \rightarrow v)$ is the rate at which a WIMP with
velocity $\omega$ scatters off a Maxwell-Boltzmann distribution of
nuclei $i$, with isotropic and velocity-independent cross
section, to a final velocity $v<\omega$~\cite{Gould:1987ju}.  In order
to account for the lack of coherence an exponential form factor
$|F_i(\omega,v)|^2$ is included~\cite{Gould:1987ir}.  We consider a
WIMP population with a local density $\rho_\chi = 0.3~{\rm GeV/cm}^3$
and a Maxwell-Boltzmann velocity distribution $f_{v_\odot}(u)$, which
as seen by an observer moving at $v_\odot$, the velocity of the Sun
with respect to the WIMPs rest frame, is given by
\begin{equation}
\label{Eq:fu}
f_{v_\odot}(u) = \sqrt{\frac{3}{2\pi}} \, \frac{u}{v_\odot \, v_d} \,
\left[e^{-\frac{3 \, (u-v_\odot)^2}{2 \, v_d^2}} - e^{-\frac{3 \,
      (u+v_\odot)^2}{2 \, v_d^2}}\right] ~,
\end{equation}
with $u$ being the WIMP velocity at infinity and $\omega^2(r) = u^2 +
v_e^2(r)$, where $v_e(r)$ is the escape velocity at a distance $r$
from the center of the Sun.  We take the values $\bar{v} =
270$~km/s for the velocity dispersion and $v_\odot = 220$~km/s for the
velocity of the Sun with respect to the WIMPs rest frame.  In the case
of neglecting either the finite temperature or the decoherence
effects, analytical solutions for the capture rate per unit volume are
known~\cite{Gould:1987ir}.  However, although the effects due to the
finite temperature of the nuclei are small, we include them in the
calculations (as well as the decoherence).

Nevertheless, Eq.~(\ref{Eq:captureweak}) is valid when the scattering
cross section is small enough so that the probability of interaction
is much smaller than 1.  However, the capture rate cannot grow
indefinitely with the cross section, for it must saturate to a maximal
value set by the geometrical cross section of the Sun (when the
probability of interaction is 1).  Using Ref.~\cite{Gould:1987ir}, the
geometrical capture rate is given by
\begin{equation}
\label{Eq:capturegeom}
C_\odot^{\rm geom} = \pi R_\odot^2 \,
\left(\frac{\rho_\chi}{m_\chi}\right) \,  
\int_0^{\infty} du \, f_{v_\odot}(u) \, \frac{\omega^2(R_\odot)}{u} = 
\pi R_\odot^2 \, \left(\frac{\rho_\chi}{m_\chi}\right)
\, \langle v \rangle_0 \,  \left(1 + \frac{3}{2} \,
\frac{v_e^2(R_\odot)}{v_d^2}\right) \xi(v_\odot) ~, 
\end{equation}
where $\langle v \rangle_0 = \sqrt{8/(3\pi)} \, v_d$ is the average
velocity in the WIMPs rest frame and the factor $\xi(v_\odot) = 0.81$
takes into account the suppression due to the motion of the Sun.  This
expression for the geometrical capture rate agrees at a level better
than the percent with that obtained in
Ref.~\cite{Bottino:2002pd}\footnote{Note that there is a typo in
  Eq.~(26) of Ref.~\cite{Bottino:2002pd}: the factor $(M_i/m_i)$
  should not be there.}.  Thus, we estimate the capture rate as
\begin{equation}
\label{Eq:capture}
C_\odot = C_\odot^{\rm weak} \, \left(1 - e^{-C_\odot^{\rm
    geom}/C_\odot^{\rm weak}}\right) ~.
\end{equation}

The annihilation rate $A_\odot$ is defined as
\begin{equation}
\label{Eq:annihilation}
A_\odot = \langle \sigma_A v\rangle \, \frac{\int_0^{R_\odot} 4\pi r^2
  dr \, n_\chi^2(r,t)}{\left(\int_0^{R_\odot} 4\pi r^2 dr \,
  n_\chi(r,t)\right)^2}  ~,
\end{equation}
where $\langle \sigma_A v\rangle$ is the thermal average of the WIMP 
annihilation cross section times the relative velocity.  In this work,
we assume an annihilation cross section typical of thermal WIMPs,
$\langle \sigma_A v\rangle = 3 \cdot 10^{-26} \, \rm{cm}^3/\rm{s}$.

Finally, analogously to the definition of the WIMP distribution, we
define the evaporation rate $E_\odot$ as
\begin{equation}
\label{Eq:evaporationgen}
E_\odot = \mathfrak{f}(Kn) \, E_{\odot, {\rm LTE}} +
\left(1-\mathfrak{f}(Kn)\right) \, E_{\odot, {\rm iso}}  ~,
\end{equation}
where, following Ref.~\cite{1990ApJ...356..302G},
\begin{equation}
\label{Eq:evaporation}
E_{\odot, \lambda} = \sum_i \int_0^{R_\odot} 4\pi r^2 \, s(r) \, dr \,
\int_0^{\infty} d\omega \, f_\odot(\omega, T(r)) \,
\int_{v_e}^{\infty} R_i^+ (\omega \rightarrow v) dv   ~, 
\end{equation}
where the rate at which a WIMP with velocity $\omega$ scatters off a
Maxwell-Boltzmann distribution of nuclei $i$, with isotropic and
velocity-independent cross section, to a final velocity $v>\omega$ is
given by $R_i^+ (\omega \rightarrow v)$~\cite{Gould:1987ju}.  We
assume that WIMPs have a truncated thermal distribution
$f_\odot(\omega, T(r))$ with a cutoff\footnote{Note that if the cutoff 
velocity is smaller than $v_e$, the evaporation rate would be
suppressed with respect to the case usually considered and that we
follow here~\cite{Gould:1987ju}.  This is a conservative approach, for
a lower evaporation rate would allow to set better limits for low WIMP
masses.} at $\omega = v_e$ where $T(r)=T_\chi$ ($T(r) = T_\odot (r)$) for
$\lambda = {\rm iso}$ ($\lambda = {\rm LTE})$.  The suppression factor
$s(r)$ accounts for the fraction of WIMPs that, after reaching the
escape velocity, would actually escape from the Sun.  We have slightly
modified the estimate of Ref.~\cite{1990ApJ...356..302G} to allow for a
smooth transition between the optically thin and thick regimes and have
defined it as
\begin{equation}
\label{Eq:suppression}
s(r) = \frac{7}{10} \, \frac{1-e^{-10 \, \tau(r)/7}}{\tau(r)} \, 
e^{-\tau(r)}  ~,
\end{equation}
where $\tau (r) = \int_r^{R_\odot} \ell(r')^{-1} \, dr'$ is the
optical depth.

In our computations we use the analytical solution for the evaporation
rate per unit volume obtained in Ref.~\cite{Gould:1987ju} and add the
suppression factor as indicated above.  This suppression is only
relevant in the optically thick regime, but indeed it results on the
evaporation mass (the minimum mass for WIMPs to be trapped in the Sun)
to {\it decrease} with the scattering cross section, which is the
opposite behavior to the one in the optically thin regime.  In other
words, the evaporation mass has a maximum at a value of the scattering
cross section around the transition between the two regimes.  This was
first noted by Ref.~\cite{1990ApJ...356..302G} and has an important
impact on our results.    

Once all the ingredients are computed, the WIMPs annihilation rate is
given by $\Gamma = A_\odot \, N_\chi^2 / 2$ and the solution of
Eq.~(\ref{Eq:evolution}) today ($t=t_\odot=4.57$~Gyr)
reads~\cite{Gaisser:1986ha, Griest:1986yu}  
\begin{equation}
\label{Eq:solution}
  \Gamma (m_\chi, \sigma_\chi) = \frac{1}{2} \, C_\odot \,
  \left(\frac{\tanh(\kappa \, t_\odot/\tau_E)}{\kappa + \frac{1}{2} \,
    E_\odot \, \tau_E \, \tanh(\kappa \, t/\tau_E)}\right)^2 \, ,
\end{equation}  
where $\tau_E = (A_\odot C_\odot)^{-1/2}$ is the equilibration time
scale in the absence of evaporation and $\kappa = (1 +
(E_\odot\tau_E/2)^2)^{1/2}$.  For a thermal annihilation cross section
and for the scattering cross sections under consideration, equilibrium
is always reached ($t_\odot >> \tau_E$, $\tanh(\kappa \,
t_\odot/\tau_E) \simeq 1$), although in our computations we keep the
exact Eq.~(\ref{Eq:solution}).

\section{MeV neutrinos from WIMPs annihilations in the Sun}
\label{sec:MeVnus}

Being produced in a very dense medium, among all the final products of
WIMPs annihilations, only neutrinos can escape.  So far, all previous
works have focused on the high-energy neutrino flux resulting from the
subsequent hadronization, fragmentation and decay of the final states
in heavy quarks, gauge bosons or $\tau^+\tau^-$ channels, and have
disregarded annihilations into $e^+e^-$, $\mu^+\mu^-$ or light quarks
because they would only produce (if any) a flux of low-energy
neutrinos from pion and muon decay at rest.  These MeV neutrinos are
the focus of this work.

The propagation of the annihilation products of WIMPs annihilations in
the Sun would produce pions that would be stopped and could
subsequently decay at rest, giving rise to a monochromatic neutrino
spectrum at 29.8~MeV.  Practically all $\pi^-$, after stopping, would
be captured in an atomic orbit and promptly the nucleus would
de-excite by emitting X-rays or transferring energy to Auger
electrons.  After that, the $\pi^-$ would get absorbed by the nucleus 
without decaying in processes of the type of $\pi^- N N \rightarrow N
N$, where $N$ represents a nucleon.  Hence, only neutrinos from
$\pi^+$ decays would contribute significantly to the low-energy
neutrino flux.  Let us note that the propagation of high-energy
$e^-/e^+$ would also produce small amounts of pions, so this could
open up the possibility to use this low-energy neutrino flux to
constrain WIMPs annihilations into $e^+e^-$.  In addition, all muons
produced in pion decays, in the leptonic decay modes of hadrons or
taus and in the case of direct annihilations into $\mu^+\mu^-$, are
stopped in the dense region where they are produced and decay at rest.
Thus, in addition to the monochromatic spectrum from $\pi^+$ decays at
rest, neutrinos from $\mu^+$ and $\mu^-$ decaying at rest would also
contribute (with a well known spectrum below $52.8$~MeV) to the final 
low-energy neutrino flux.  Hence, the sources of the flux of neutrinos
studied here are
\begin{eqnarray}
\label{Eq:spectra}
\pi^+ & \rightarrow & \mu^+ + \nu_\mu ~,\nonumber \\
\pi^- & \rightarrow & \mu^- + \bar{\nu}_\mu \hspace{5mm} {\rm \it
  (negligible \, \, contribution)}~,\nonumber \\
\mu^+ & \rightarrow & e^+ + \nu_e + \bar{\nu}_\mu ~, \\
\mu^- & \rightarrow & e^- + \bar{\nu}_e + \nu_\mu ~. \nonumber
\end{eqnarray}

\begin{figure}[t]
\begin{center}
\includegraphics[width=0.45\linewidth]{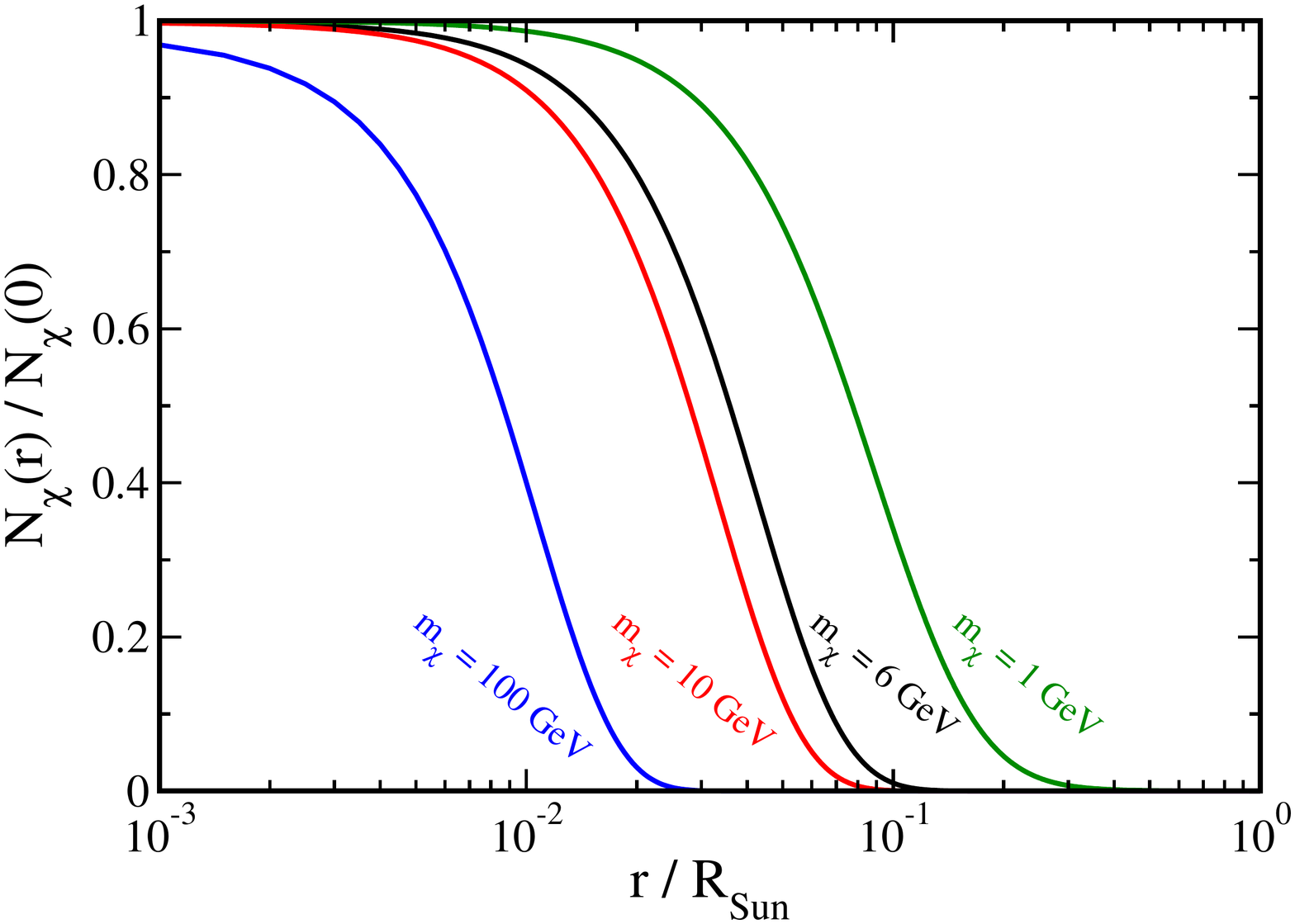}
\hspace{5mm}
\includegraphics[width=0.45\linewidth]{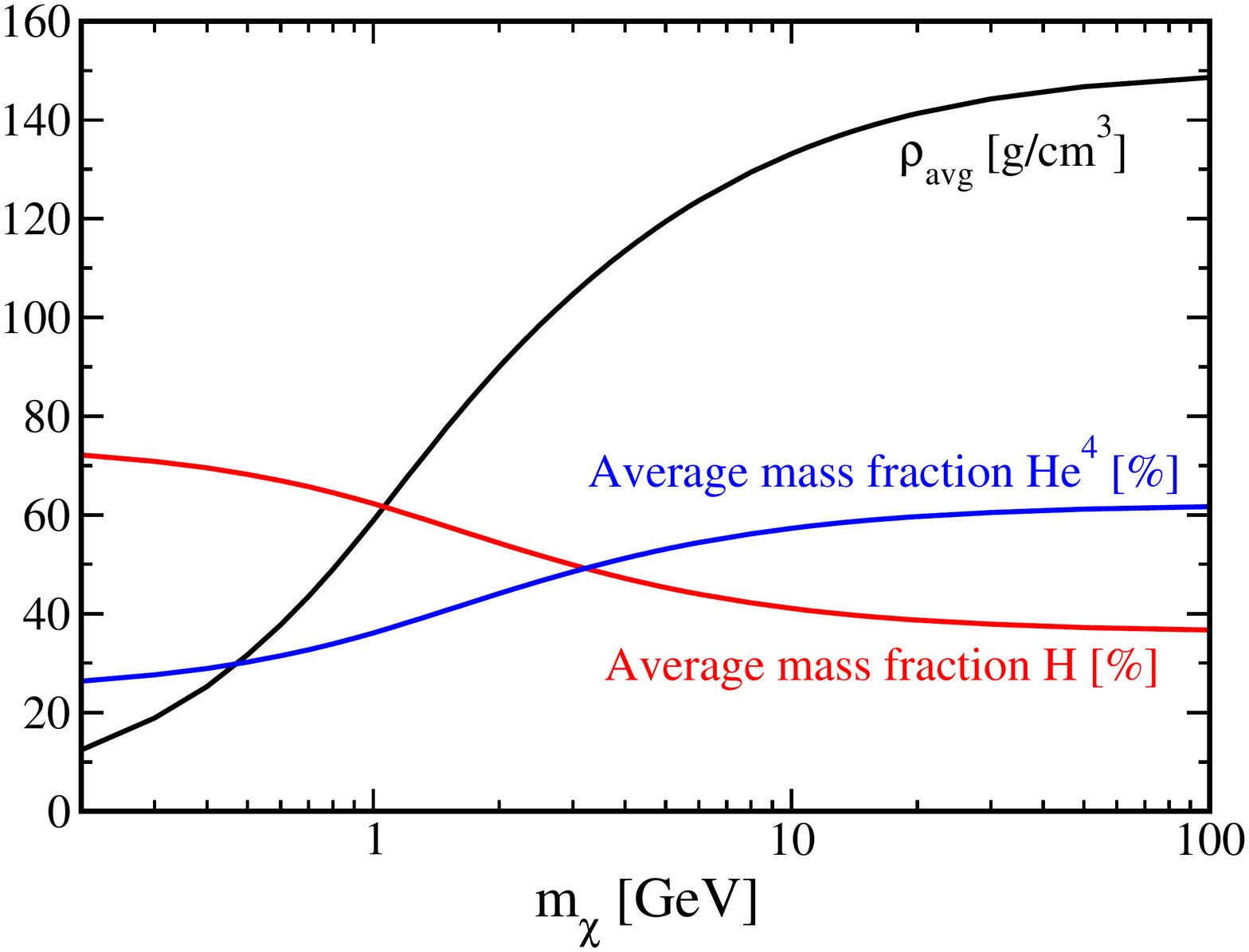}
\caption{\sl \footnotesize \textbf{\textit{Left panel: Distribution of
      the number density of WIMPs in the Sun}} (normalized to the
      density at the center of the Sun), $n_\chi(r,t)/n_\chi(0,t)$, as a
  function of the distance to the center of the Sun for four WIMP
  masses.  \textbf{\textit{Right panel: Weighted density and
      composition of the Sun}}, according to the
  SSM~\cite{Asplund:2009fu, Serenelli:2011py} and to the distribution
  of WIMPs (left panel), as a function of the WIMP mass.  We only show
  the two main elements, He$^4$ and H.  Here, we have assumed a
  spin-dependent cross section, $\sigma_{\rm p}^{\rm SD} =
  10^{-40}$~cm$^2$, although the results are almost the same for any
  other case.} 
\label{fig:fig1}
\end{center}
\end{figure}

The shape of the spectra of these neutrinos is well known and, in
order to calculate the relative contributions of each type of neutrino
spectrum to the final neutrino spectrum, we simulate all the particles
propagation with GEANT4~\cite{Agostinelli:2002hh, Allison:2006ve}.  To
determine for each WIMP mass the average density and composition of
the medium where the products of WIMPs annihilations propagate, we use
the SSM~\cite{Asplund:2009fu, Serenelli:2011py} and the WIMP
distribution in the Sun (as a function of the WIMP mass), given in
Eq.~(\ref{Eq:WIMPdist}), to compute it.  Thus, for each WIMP mass,
we obtain the average density and solar composition of the region
where WIMPs accumulate and annihilate.  In Fig.~\ref{fig:fig1} we show
the radial distribution of the number density of WIMPs in the Sun
normalized to its value at the solar center,
$n_\chi (r, t)/n_\chi (0, t)$, (left panel) and the weighted density and
composition of the Sun in the WIMPs environment (right panel).

For the simulations, we proceed as follows.  For the case of WIMPs
annihilating into a pair of leptons we inject the two leptons with
energies equal to the WIMP mass directly into GEANT4 and let them
propagate and decay. In the case of WIMPs annihilations into quarks,
we use PYTHIA 6.4~\cite{Sjostrand:2006za} to hadronize and fragment
the initial quarks and do not let decay any of the final particles
that are produced.  Then, we inject into GEANT4 the full spectrum of
all the produced particles and simulate their propagation in the Sun.
Finally, we count the number of $\pi^+$, $\pi^-$, $\mu^+$ and $\mu^-$  
that decay at rest (we also count all $\pi^-$, which are not captured
and decay at rest, although their number is negligible).  These
numbers represent the relative weights for each of the types of
neutrino and antineutrino spectra (four from muon decay and two from 
pion decay) indicated in Eq.~(\ref{Eq:spectra}), which allow us to
compute the initial electron and muon neutrino and antineutrino fluxes
at the source. 

\begin{figure}[t]
\begin{center}
\includegraphics[width=0.32\linewidth]{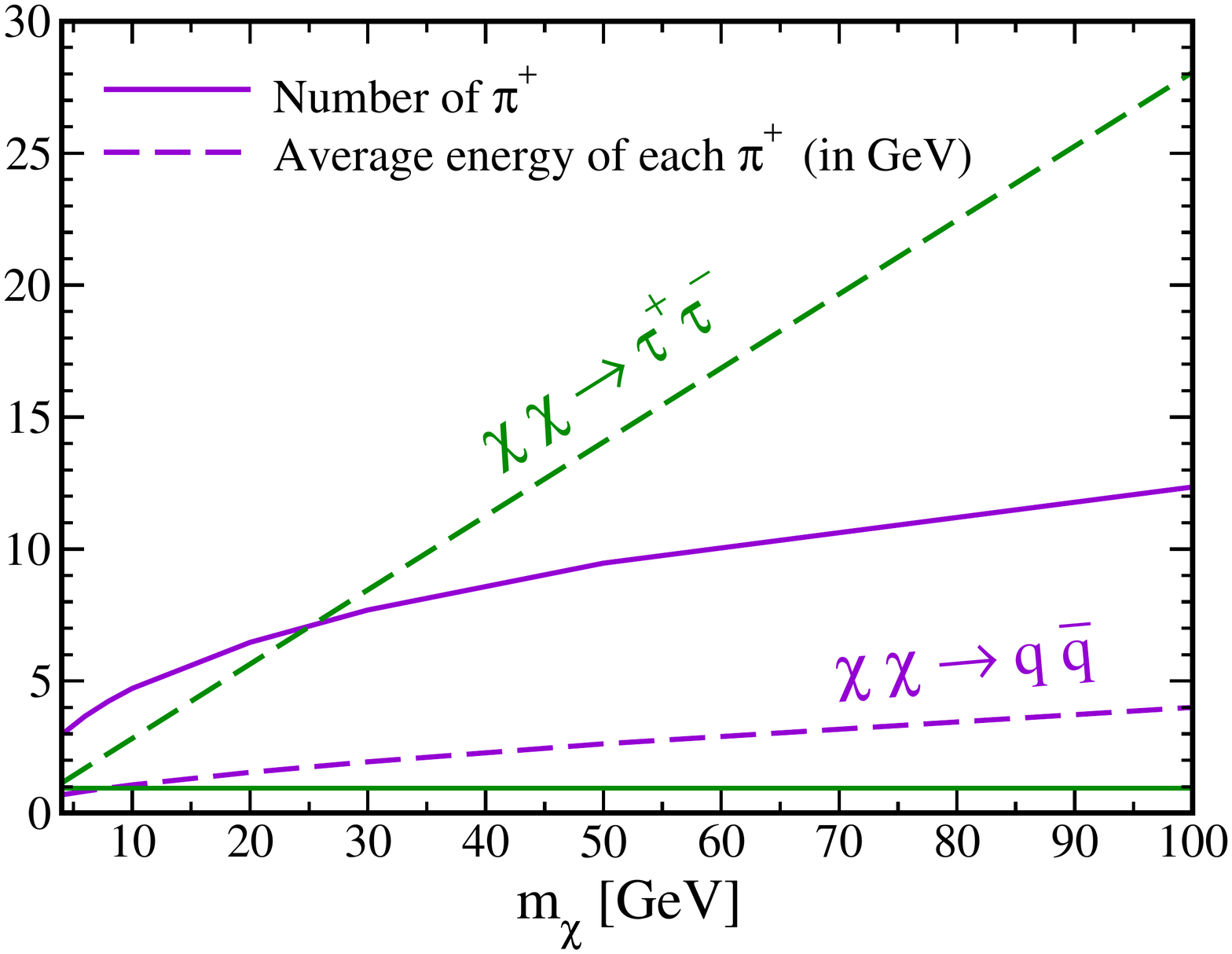}
\hspace{0.5mm}
\includegraphics[width=0.32\linewidth]{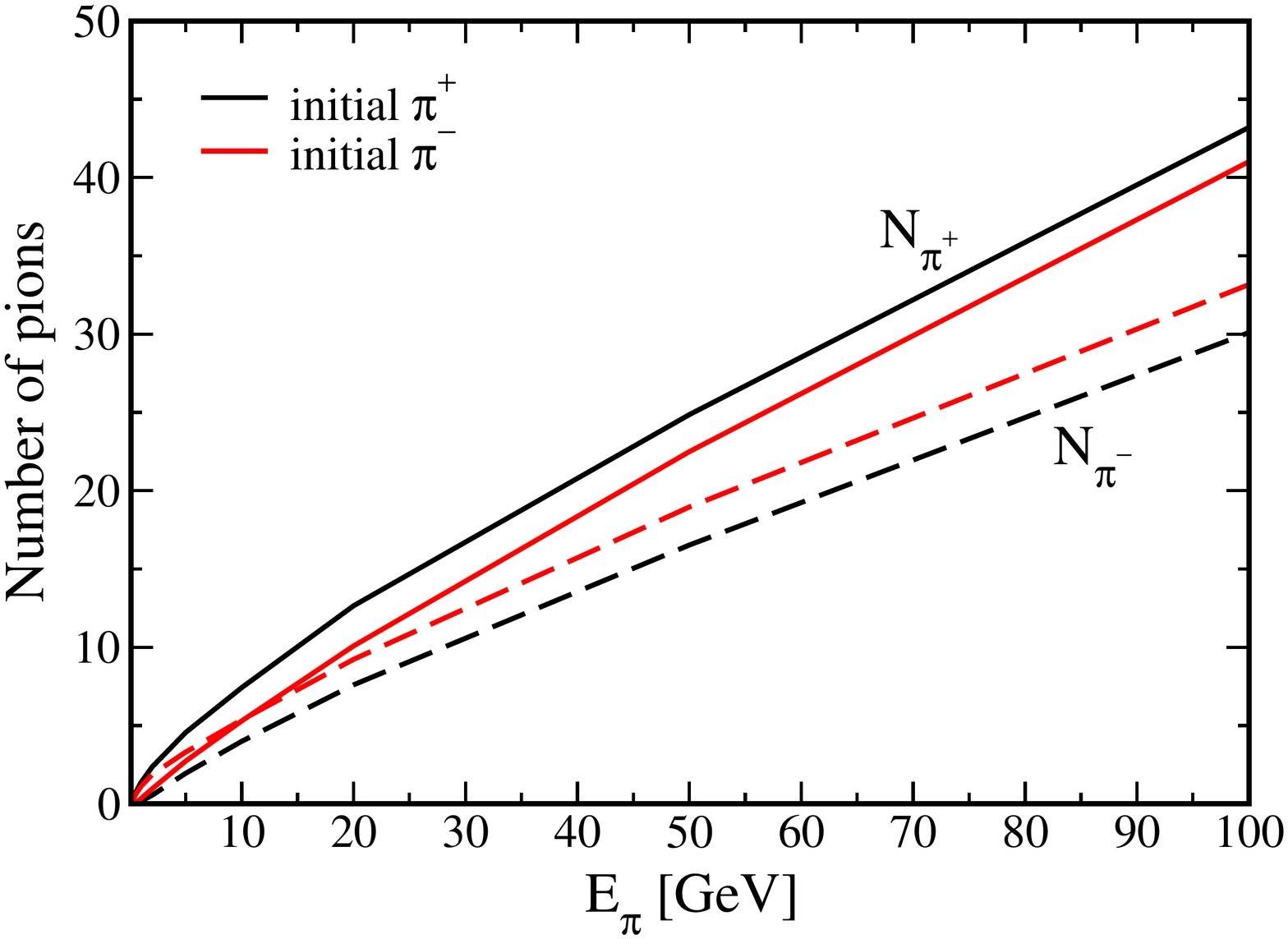}
\hspace{0.5mm}
\includegraphics[width=0.32\linewidth]{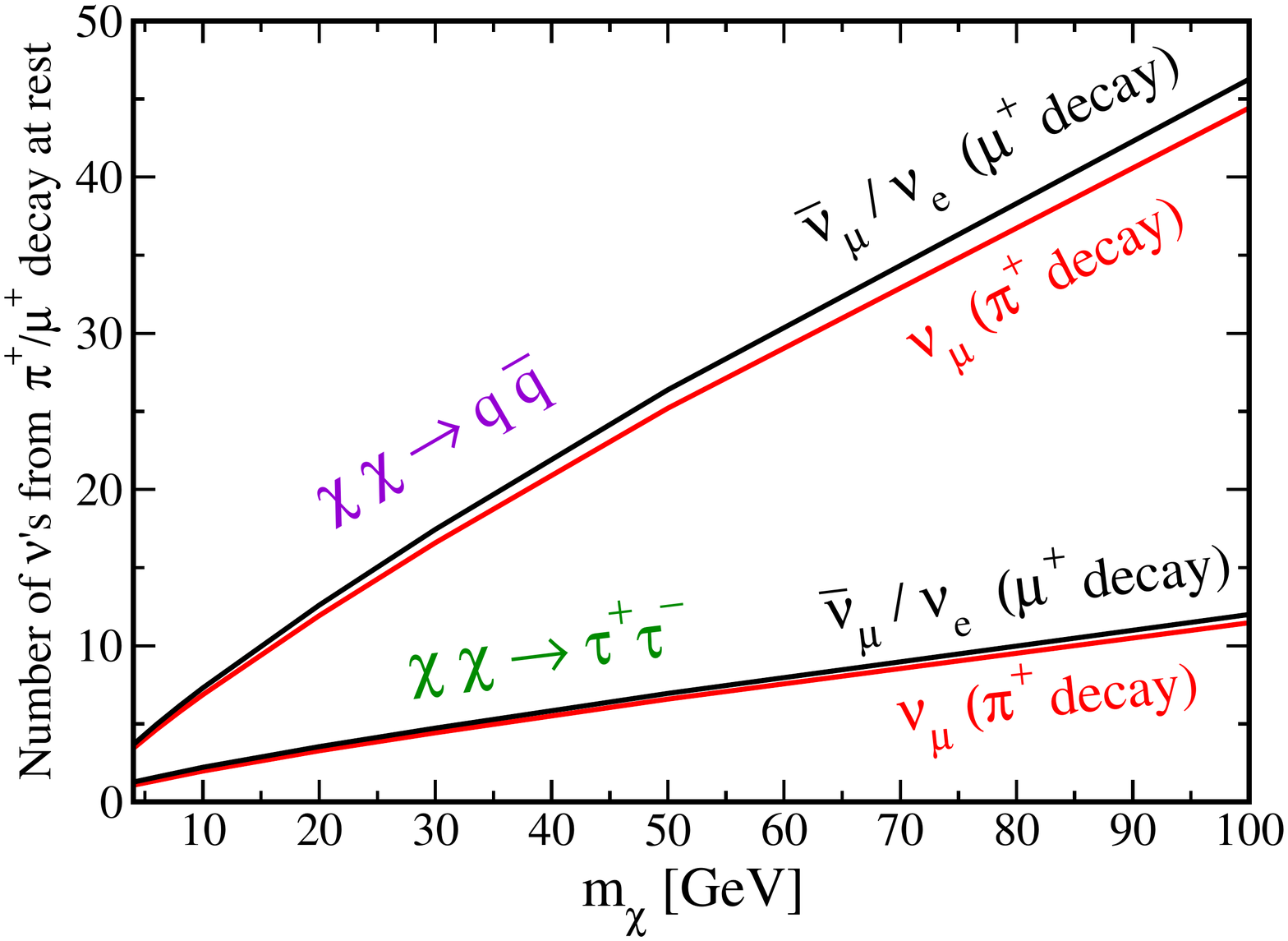}
\caption{\sl \footnotesize \textbf{\textit{Left panel: Number of
      $\pi^+$ (solid lines) and their average energy (dashed lines)}}
      just after WIMPs annihilations (before propagation) as a
      function of the WIMP mass.  \textbf{\textit{Middle panel: Number
          of $\pi^+$ (solid lines) and $\pi^-$ (dashed lines) produced
          after the propagation of one $\pi^+$ (black lines) or one
          $\pi^-$ (red lines)}} as a function of the energy of the
      initial pion.  \textbf{\textit{Right panel: Number of neutrinos
          and antineutrinos from $\pi^+$ and $\mu^+$ decay at rest}}
      as a function of the WIMP mass.  In the three panels, we show the
      results for two annihilation channels, $\chi \chi \rightarrow q
      \bar{q}$ and $\chi \chi \rightarrow \tau^+ \tau^-$.  We have
      considered an average density and composition of the region where
      WIMPs annihilation occur corresponding to $m_\chi = 100$~GeV.}    
\label{fig:fig2}
\end{center}
\end{figure}

The final number of neutrinos and antineutrinos from pion and muon
decay at rest can be understood from Fig.~\ref{fig:fig2} where we
show different results for WIMPs annihilations into light quarks and tau
leptons.  In the left panel, we show the number of $\pi^+$ (solid
lines) and their average energy (dashed lines) just after WIMPs
annihilations (before propagation) as a function of the WIMP mass; in
the middle panel, the number of $\pi^+$ (solid lines) and $\pi^-$
(dashed lines) produced after the propagation of one $\pi^+$ (black
lines) or one $\pi^-$ (red lines) as a function of the energy of the
initial pion; and in the right panel, the number of neutrinos and
antineutrinos from $\pi^+$ and $\mu^+$ decay at rest as a function of
the WIMP mass.  From the left panel we see that, whereas the number of
$\pi^+$ produced (before propagation of the products of annihilation)
in the case of annihilations into light quarks grows with the WIMP
mass and in the case of annihilations into tau leptons it remains
approximately constant, the average energy of these pions increases
faster with the WIMP mass in the latter case.  Overall, convolving
this result with that in the middle panel, this behavior implies a
faster increase on the final number of neutrinos for WIMPs
annihilations into light quarks than into tau leptons (see right
panel), and hence better limits in the former case, as discussed below.

These neutrinos would then propagate from the Sun to the Earth and
would be detected via the charged-current interactions of those
arriving at the detector in the electron flavor (see below).  In order
to calculate the electron neutrino and antineutrino fluxes at the Earth,
one should take into account that neutrinos mix.  For the energies of
interest (above 10~MeV) neutrinos propagate adiabatically in the Sun
and at the region where they are produced matter effects are dominant,
so $\nu_e$ ($\bar{\nu}_e$) exit the Sun as almost purely $\nu_2$
($\nu_1$) and $\nu_\mu$ ($\bar{\nu}_\mu$) almost as an equal mixture
of $\nu_1$ ($\nu_2$) and $\nu_3$.  Hence, the probabilities for
neutrinos to arrive at the Earth in the electron flavor are
\begin{eqnarray}
  P(\nu_\mu \rightarrow \nu_e) & =  & \sin^2\theta_{13} \,
  \cos^2\theta_{13} \, \sin^2\theta_{12} \, (1 + \sin^2\theta_{12}) +
  \cos^2\theta_{23} \, \cos^2\theta_{12} \, \cos^2\theta_{13} \simeq
  0.35 ~, \nonumber \\ 
  P(\bar{\nu}_\mu \rightarrow \bar{\nu}_e) & = &  \sin^2\theta_{13} \,
  \cos^2\theta_{13} \, \sin^2\theta_{12} \, (1 + \cos^2\theta_{12}) +
  \cos^2\theta_{23} \, \sin^2\theta_{12} \, \cos^2\theta_{13} \simeq
  0.17 ~, \nonumber \\ 
  P(\nu_e \rightarrow \nu_e) & = & \sin^2\theta_{12} \,
  \cos^2\theta_{13} + \sin^4\theta_{13} \simeq 0.31 ~, \nonumber  \\
  P(\bar{\nu}_e \rightarrow \bar{\nu}_e) & = & \cos^2\theta_{12} \,
  \cos^2\theta_{13} + \sin^4\theta_{13}  \simeq 0.66 ~.
\end{eqnarray}
The final electron neutrino and antineutrino spectra at the detector
are obtained by combining the initial fluxes with the above
probabilities\footnote{We neglect the correction due to the Earth-matter
effect~\cite{Lunardini:2001pb, Peres:2009xe}, which after averaging
over all possible trajectories we expect it to be at the percent
level.  Moreover, the SK collaboration did not take this into account
when simulating the backgrounds and we use their results in our
analysis.}.  We use the values of the mixing angles (for normal
hierarchy) from Ref.~\cite{Tortola:2012te}.

\section{Detection of MeV neutrinos with SK}
\label{sec:Detection}

SK is a water \v{C}erenkov detector with a fiducial volume of
22.5~ktons ($1.5 \cdot 10^{33}$ free protons and $7.5 \cdot 10^{32}$
oxygen nuclei).  For energies below $52.8$~MeV, the best present data
come from the search for the DSNB~\cite{Bays:2011si, Baysthesis},
which is split into three phases: SK-I ($t_I = 1497$~days), SK-II
($t_{II} = 794$~days) and SK-III ($t_{III} = 562$~days).  The recent
analysis~\cite{Bays:2011si, Baysthesis} has substantially improved
over the previous one~\cite{Malek:2002ns, Malekthesis}, with better
efficiencies, lower energy thresholds and almost twice as much
statistics.

The signal at SK is the detection of the positrons (electrons) produced
in $\bar{\nu}_e$ ($\nu_e$) charged-current interactions in the
detector below $\sim$~100~MeV.  At these energies, the inverse
beta-decay cross section ($\bar{\nu}_e \, p \rightarrow n \, e^+$) is
about two orders of magnitude larger than the $\nu-e$ elastic scattering
cross section.  Below $\sim$~80~MeV, this is the dominant interaction of
$\bar{\nu}_e$.  Although the WIMP signal discussed in this work is
below this energy, we have also taken into account the interactions of 
$\nu_e$ and $\bar{\nu}_e$ off oxygen nuclei, which give non-negligible
contributions to the lowest energy bins. 

The low energy threshold in the analysis is determined by the ability
to remove the radioactive spallation caused by cosmic-ray muons
hitting an oxygen nucleus.  An improved spallation cut has allowed to
reduce the energy threshold used in the previous analysis down to 16~MeV
(17.5~MeV) for SK-I/III (SK-II).  In addition to this, a number of
other cuts were performed, as noise reduction, fiducial volume, solar
angle, incoming event, decay electron, pion, \v{C}erenkov angle and
other cuts~\cite{Bays:2011si, Baysthesis}.  The maximum energy in the
recent SK analysis is 88~MeV, which is also higher than in the
previous one.

In this energy range, the two dominant backgrounds which remain after
the cuts are the atmospheric $\nu_e$ and $\bar{\nu}_e$ background and,
mainly, the Michel positrons (electrons) from the decays of low
energy muons, produced by atmospheric $\bar{\nu}_\mu$ ($\nu_\mu$) with
typical energies of about $\sim$200~MeV, which are below detection
threshold, the so-called invisible muons\footnote{If the momentum of the
produced muon is below $\sim$120~MeV, then the muon is below the
threshold for emitting \v{C}erenkov radiation in water.}.  These muons
are slowed down rapidly and subsequently decay, mimicking the signal
from $\nu_e$ or $\bar{\nu}_e$, but with a spectrum whose shape is very
well known.

In addition to these two backgrounds, in the new SK analysis two extra
sources of background were considered: neutral current (NC) elastic
events, which give rise to de-excitation gammas or produce other
reactions, and low energy muons and pions misidentified as
electrons/positrons.

In the new SK analysis three \v{C}erenkov angle regions are defined:
$20^\circ-30^\circ$ (the `low angle' or `$\mu/\pi$' region),
$38^\circ-50^\circ$ (the `signal' region) and $78^\circ-90^\circ$ (the
`high angle' or `NC elastic' region).   Inverse beta-decay positrons
in the data sample (with energies above 16~MeV) are highly
relativistic and have a \v{C}erenkov angle of around $42^\circ$.  On
the other hand, low energy muons and pions travel more slowly and emit
light with a smaller \v{C}erenkov angle.  In addition, some other
events with more isotropic nature can have higher \v{C}erenkov angles,
such as events with multiple gammas.

\section{Analysis}
\label{sec:Analysis}

In this work, we follow some parts of the analysis performed in
Refs.~\cite{PalomaresRuiz:2007eu, PalomaresRuiz:2007ry}, although we
update some aspects following the new SK analysis~\cite{Bays:2011si,
  Baysthesis}, as we explain below.

\begin{figure}[t]
\begin{center}
\includegraphics[width=0.45\linewidth]{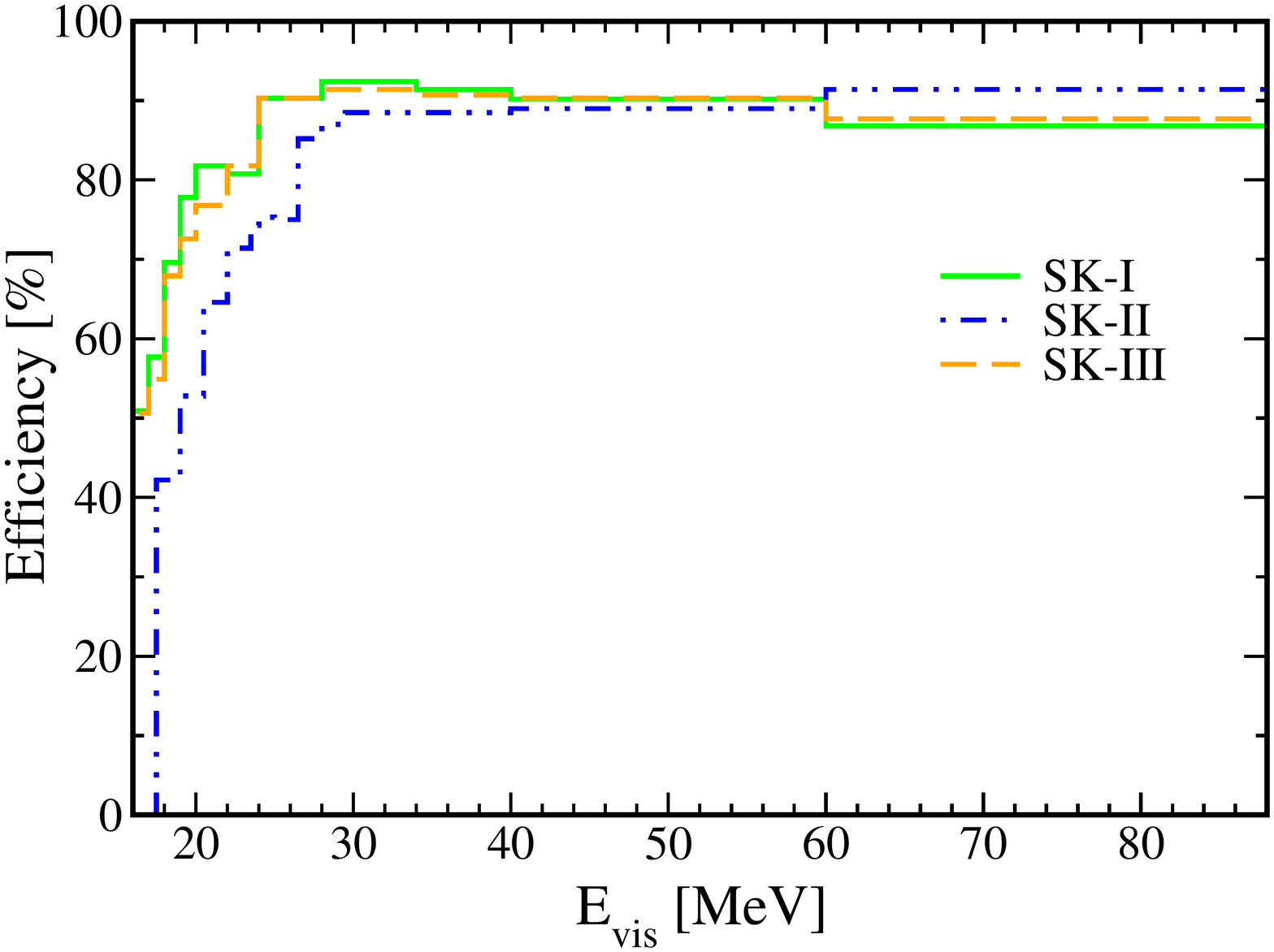}
\hspace{5mm}
\includegraphics[width=0.45\linewidth]{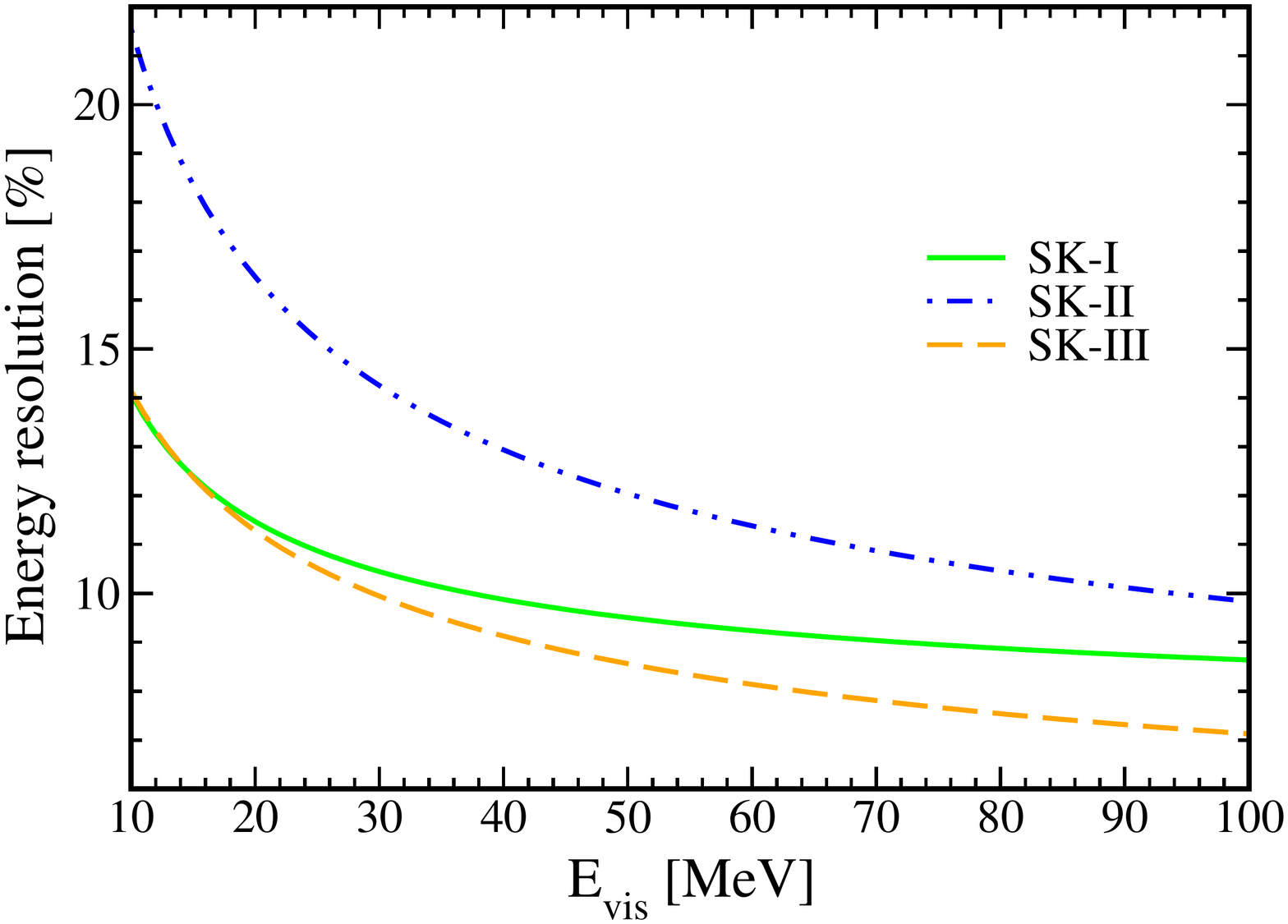}
\caption{\sl \footnotesize
  \textbf{\textit{Efficiency~\cite{Bays:2011si, Baysthesis} (left
      panel) and energy resolution~\cite{Hosaka:2005um, Cravens:2008aa,
        Abe:2010hy} (right panel) of each SK phase}} as a function of
  the visible (detected positron or electron) energy, $E_{\rm vis}$.}
\label{fig:fig3}
\end{center}
\end{figure}

We have considered both the interactions of $\bar{\nu}_e$ off free
protons and the interactions of $\nu_e$ and $\bar{\nu}_e$ off bound
nucleons.  At very low energies, the inverse beta-decay reaction
relates the energy of the outgoing positron to that of the incoming
$\bar{\nu}_e$, such that $E_{e^+} \simeq E_{\bar{\nu}} -
1.3$~MeV.  However, at higher energies, corrections of the order of
${\cal O} (E_{\bar{\nu}}/\rm{M}_{\rm{N}})$, where $\rm{M}_{\rm{N}}$ is the
nucleon mass, become important and the spread in positron energy is,
to first order, given by $\Delta E_{e^+} \approx 2
E_{\bar{\nu}}^2/\rm{M}_{\rm{N}}$.  For the inverse beta-decay reaction
we use the full cross section~\cite{Strumia:2003zx, Vogel:1999zy} and
for the interactions off bound nucleons, we consider a relativistic
Fermi gas model~\cite{Smith:1972xh} with a Fermi surface momentum of
225~MeV and a binding energy of 27~MeV.  For each of the three SK
phases, we have used the corresponding energy-dependent
efficiencies~\cite{Bays:2011si, Baysthesis} and a Gaussian energy
resolution function~\cite{Hosaka:2005um, Cravens:2008aa, Abe:2010hy}
of width $\sigma (E_e)$, $R(E_e,E_{\rm vis})$ (shown in
Fig.~\ref{fig:fig3}), with $E_e$ and $E_{\rm vis}$ being the incoming
and detected electron/positron energy, respectively.

\begin{figure}[t]
\begin{center}
\includegraphics[width=0.45\linewidth]{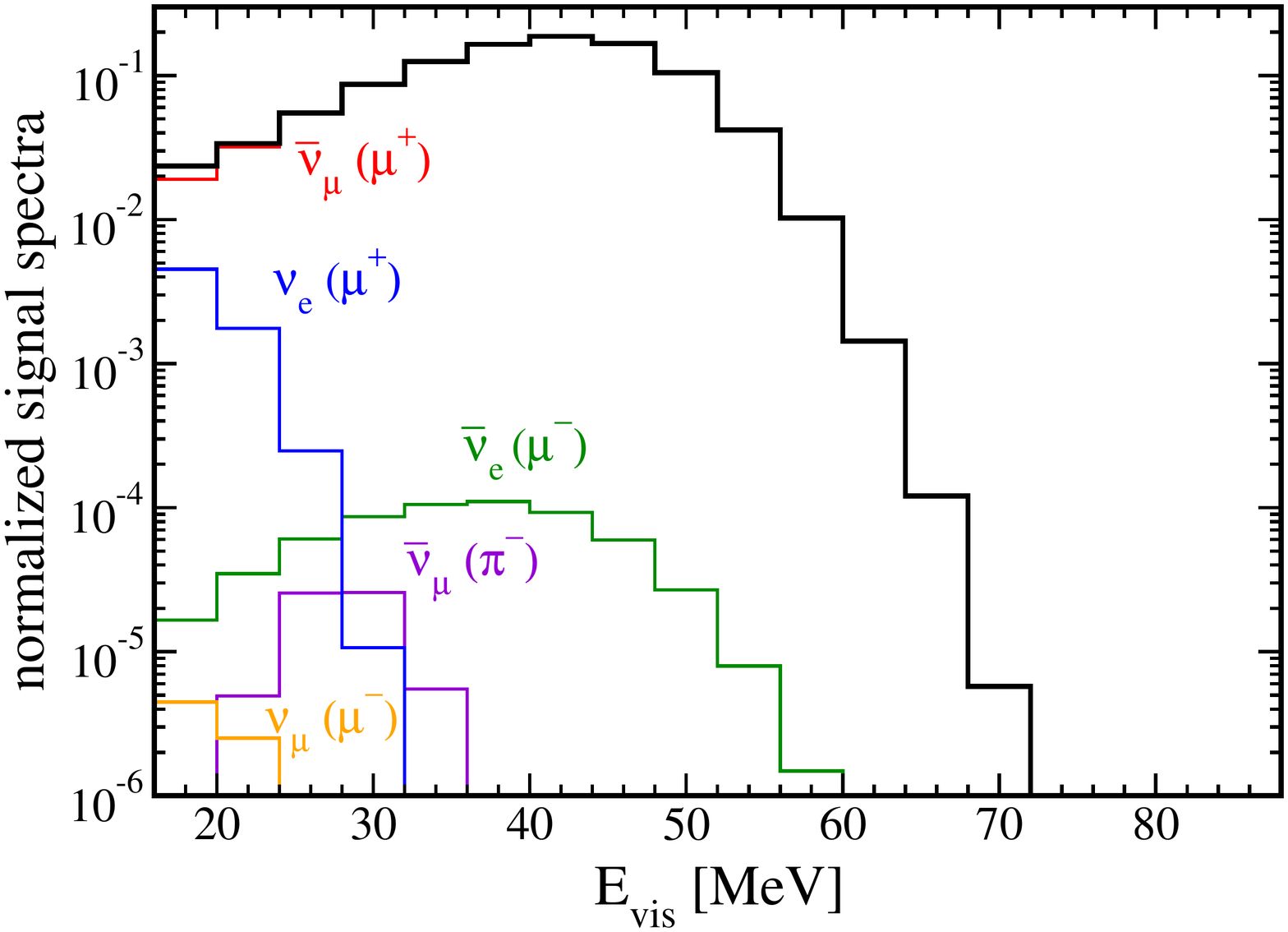}
\hspace{5mm}
\includegraphics[width=0.45\linewidth]{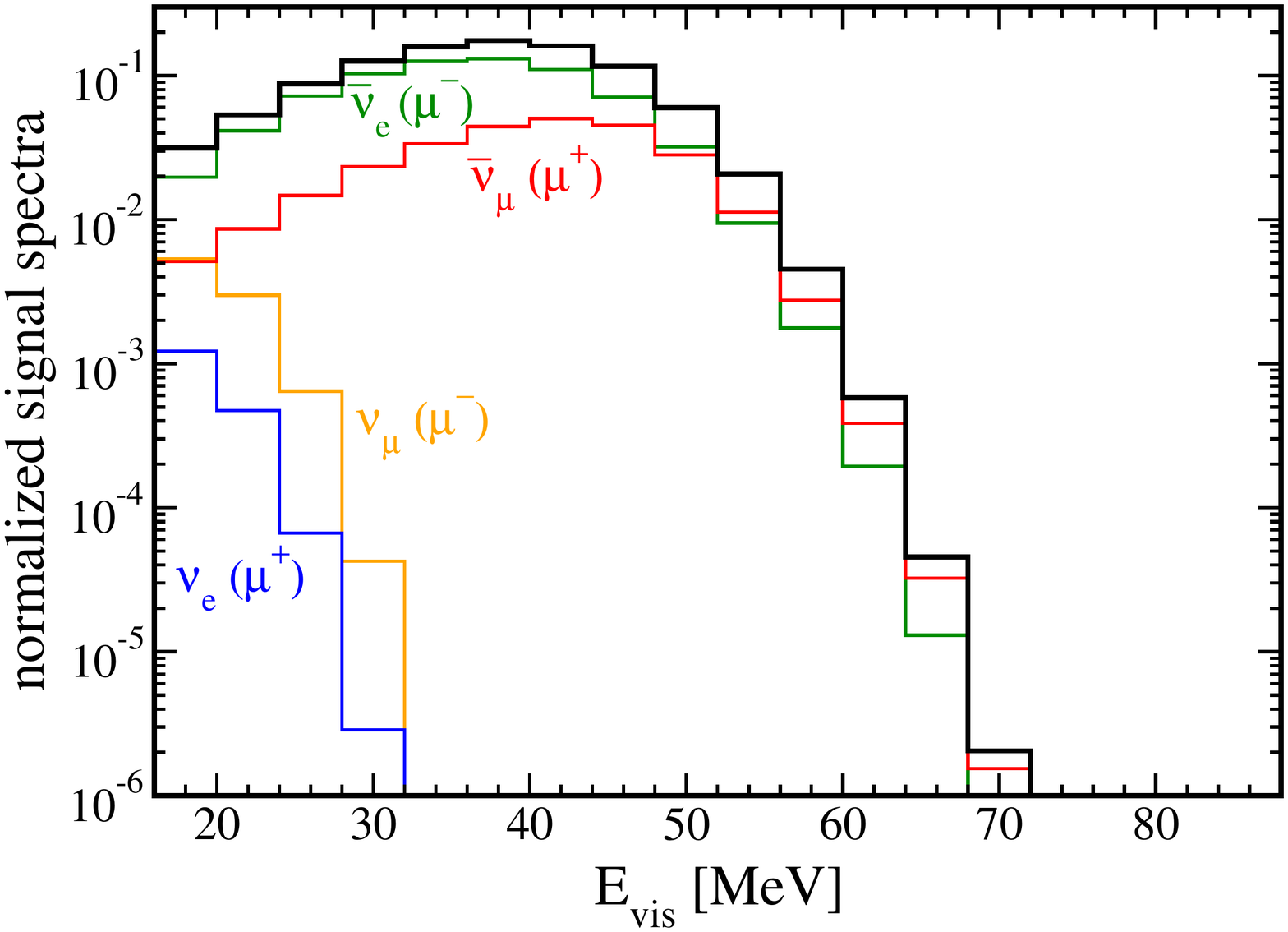}
\caption{\sl \footnotesize \textbf{\textit{Normalized signal spectra}}
  in the region of observation for WIMPs annihilations into light
  quarks (left panel) and $\mu^+\mu^-$ (right panel), for $m_\chi =
  6$~GeV and SK-I.  The colored lines represent the different
  contributions to the final neutrino spectra from pion (in the left
  panel) and muon decay (in both panels) at rest and the thick black
  line represents the total spectrum.
}   
\label{fig:fig4}
\end{center}
\end{figure}

The expected fraction of the low-energy neutrino signal from WIMPs
annihilations in the Sun in the visible electron/positron energy
interval $E_{\rm vis} = [E_l, E_{l+1}]$ is given by
\begin{eqnarray}
  A_l  =   A_s \, \int dE_e \, dE_\nu \, G_l (E_e) & \times & \left[
    \left( \frac{d\sigma_{\rm f}^{\bar{\nu}_e}}{dE_e} 
    (E_{\bar{\nu}_e},E_e) + \frac{1}{2} \, 
    \frac{d\sigma_{\rm b}^{\bar{\nu}_e}}{dE_e} (E_{\bar{\nu}_e},E_e)
    \right) \, \frac{d\Phi^{\bar{\nu}_e}}{dE_{\nu_e}} (E_{\bar{\nu}_e})
    \right. \nonumber \\
    & & \left. +
    \frac{1}{2} \, \frac{d\sigma_{\rm b}^{{\nu}_e}}{dE_e}
    (E_\nu,E_e)  \, \frac{d\Phi^{\nu_e}}{dE_{\nu_e}} (E_{\nu_e}) \right]
     ~,
\end{eqnarray} 
with $E_1 = 16$~MeV and $E_{l+1} - E_l = 4$~MeV. $A_s$ is a
normalization constant so that $\sum A_l = 1$.  The neutrino cross
sections off free nucleons and off nuclei (bound nucleons) are given
by $\sigma_{\rm f}$ and $\sigma_{\rm b}$, respectively, and the factor
$1/2$ is due to water having twice as many free protons as oxygen
nuclei.  The effects of the energy resolution function are embedded in
$G_l (E_e) = \int_{E_l}^{E_{l+1}} \, \epsilon (E_{\rm vis}) \,
R(E_e,E_{\rm vis}) \, dE_{\rm vis}$, with $\epsilon (E_{\rm vis})$ the
efficiency in that energy bin.  The fluxes of electron neutrinos and
antineutrinos at the detector are given by
\begin{eqnarray}
\frac{d\Phi^{\nu_e}}{dE_{\nu_e}} (E_{\nu_e}) & = & \frac{1}{4 \pi
  d_\odot^2} \, \Gamma (m_\chi, \sigma_\chi) \, \left( P(\nu_e
\rightarrow \nu_e) \, \frac{dF}{dE_{\nu_e}} (E_{\nu_e}) + P(\nu_\mu
\rightarrow \nu_e) \, \frac{dF}{dE_{\nu_\mu}} (E_{\nu_\mu}) \right)
\nonumber \\ 
\frac{d\Phi^{\bar{\nu}_e}}{dE_{\bar{\nu}_e}} (E_{\bar{\nu}_e}) & = &
\frac{1}{4 \pi d_\odot^2} \, \Gamma (m_\chi, \sigma_\chi) \, \left(
P(\bar{\nu}_e \rightarrow \bar{\nu}_e) \, 
\frac{dF}{dE_{\bar{\nu}_e}} (E_{\bar{\nu}_e}) + P(\bar{\nu}_\mu
\rightarrow \bar{\nu}_e) \, \frac{dF}{dE_{\bar{\nu}_\mu}}
(E_{\bar{\nu}_\mu}) \right)  ~,
\end{eqnarray}
where $d_\odot$ is the average distance Sun-Earth and $dF/dE_{\nu_e}$ and
$dF/dE_{\bar{\nu}_e}$ ($dF/dE_{\nu_\mu}$ and $dF/dE_{\bar{\nu}_\mu}$)
are the electron (muon) neutrino and antineutrino spectra per WIMPs
annihilation in the Sun.  We show in Fig.~\ref{fig:fig4} the
normalized signal spectra of the different contributions to the final
neutrino spectra in the interval $E_{\rm vis} = [16-88]$~MeV for the
case of WIMPs annihilations into light quarks (left panel) and
$\mu^+\mu^-$ (right panel), for $m_\chi = 6$~GeV and SK-I.

In order to obtain the upper limit on the WIMP-nucleon scattering
cross section, we use the recent data reported by the SK
collaboration~\cite{Bays:2011si, Baysthesis}.  We consider the four
types of background described above and fit the data in the three
\v{C}erenkov angle regions defined in the SK
analysis~\cite{Bays:2011si, Baysthesis}.  In order to do so, we use
the probability distribution functions (PDF) provided in
Ref.~\cite{Baysthesis}, which include the relative normalizations
among the three \v{C}erenkov regions for each background and each SK 
phase.  In the analysis we leave the normalizations of the four
backgrounds and the signal free in each \v{C}erenkov region, but with
the relative normalizations among regions kept fixed.  We fit the
total number of events of each type, $\alpha$ (WIMP signal), $\beta$
(invisible muons background), $\gamma$ (atmospheric $\nu_e$
background), $\delta$ (NC elastic background) and $\eta$ ($\mu/\pi$
background).  We consider 18 4-MeV bins (in the interval 16-88~MeV)
and perform an extended maximum likelihood fit.  We obtain the best
fit as the combination of parameters that maximizes the likelihood,
defined as
\begin{equation}
\mathcal{L} = e^{-(\alpha + \beta + \gamma + \delta + \eta)} \,
\prod_{a=1}^{3} \, \prod_{l=1}^{18} \frac{\left[(\alpha \cdot A_l^a) +
  (\beta \cdot B_l^a) + (\gamma \cdot C_l^a) + (\delta \cdot D_l^a) +
  (\eta \cdot E_l^a) \right]^{N_l^a}}{N_l^a!} ~,
\end{equation}
where the product $a$ is over the three \v{C}erenkov regions, the
product $l$ is over all energy bins, $N_l^a$ is the number of
detected events in the $l$-th bin in region $a$, and $A_l^a$, $B_l^a$,
$C_l^a$, $D_l^a$ and $E_l^a$ are the fractions (so that for each case
the total is normalized to 1 over the three regions) of the WIMPs
annihilation signal, Michel positrons and electrons from muon decay,
atmospheric $\nu_e$- and $\bar{\nu}_e$-induced spectra, NC elastic
events and misidentified muons and pions, that are in the $l$-th bin
and in the $a$ \v{C}erenkov region, respectively.  The fractions
$A_l^{\rm signal}$ in the signal region are calculated using the
$\bar{\nu}_e$ and $\nu_e$ low-energy fluxes from WIMP annihilations in
the Sun as described above\footnote{As the spectrum of the WIMP signal
  is similar to that of the invisible muons background, we have
  estimated the signal in the three \v{C}erenkov angle regions as
  $A_l^a \propto A_l^{\rm signal} \times B_l^a/B_l^{\rm signal}$ and then
  properly normalized to 1 over the three regions. Nevertheless, we
  get the same results if we assume that the events from WIMPs
  annihilations are only in the signal region.}.  We have reproduced
the PDFs in the signal region for the two main backgrounds and our
results are in perfect agreement with the SK results.  In order to
reproduce $B_l^{\rm signal}$ and $C_l^{\rm signal}$ we have used the
atmospheric neutrino flux calculation with
FLUKA~\cite{Battistoni:2002ew, Battistoni:2005pd}, and for $B_l^{\rm
  signal}$ we have taken into account that 18.4\% of the $\mu^-$ are
stopped in water, so the spectrum of their decay electron gets
distorted.

\begin{figure}[t]
\begin{center}
\includegraphics[width=0.45\linewidth]{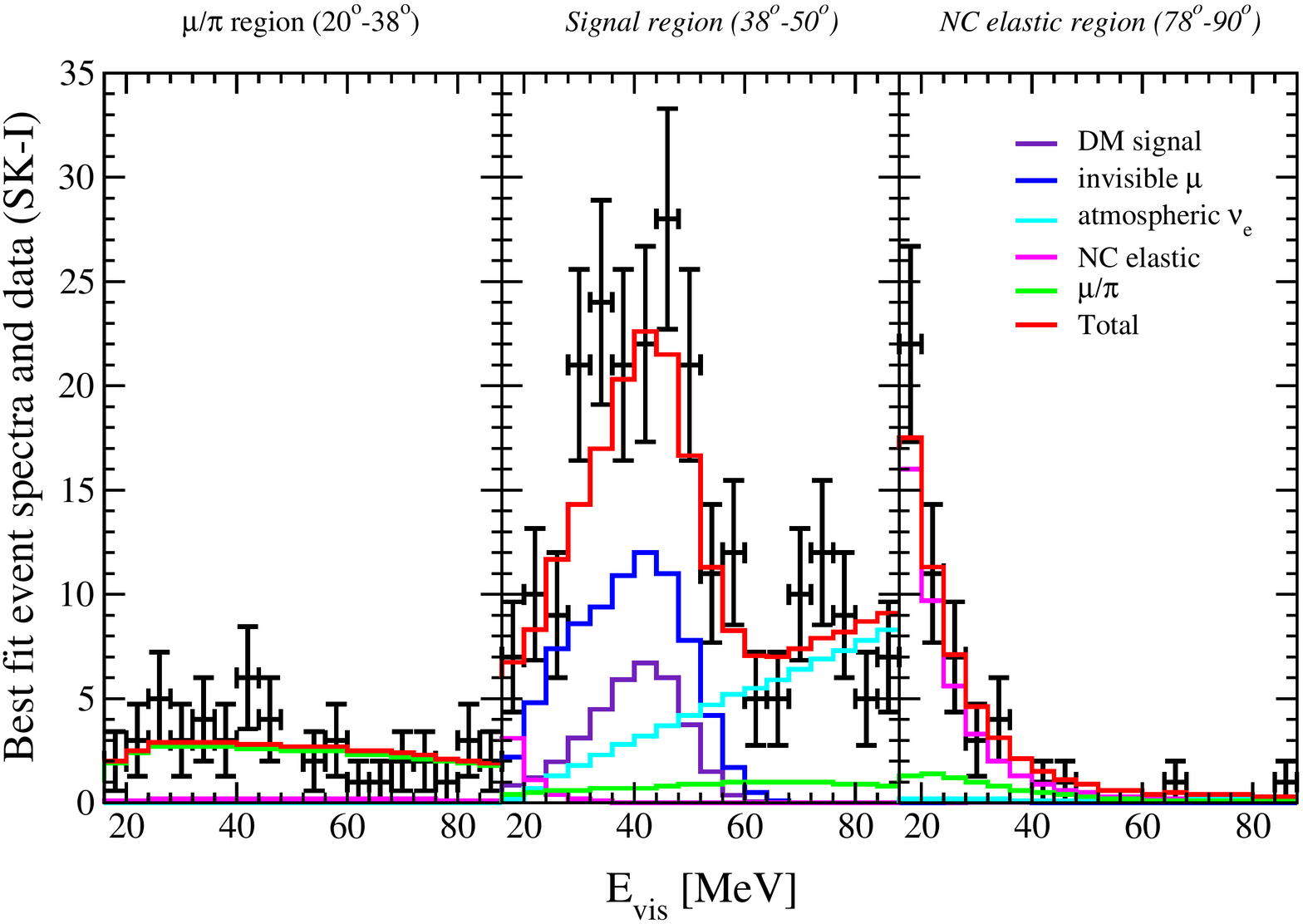}
\hspace{5mm}
\includegraphics[width=0.45\linewidth]{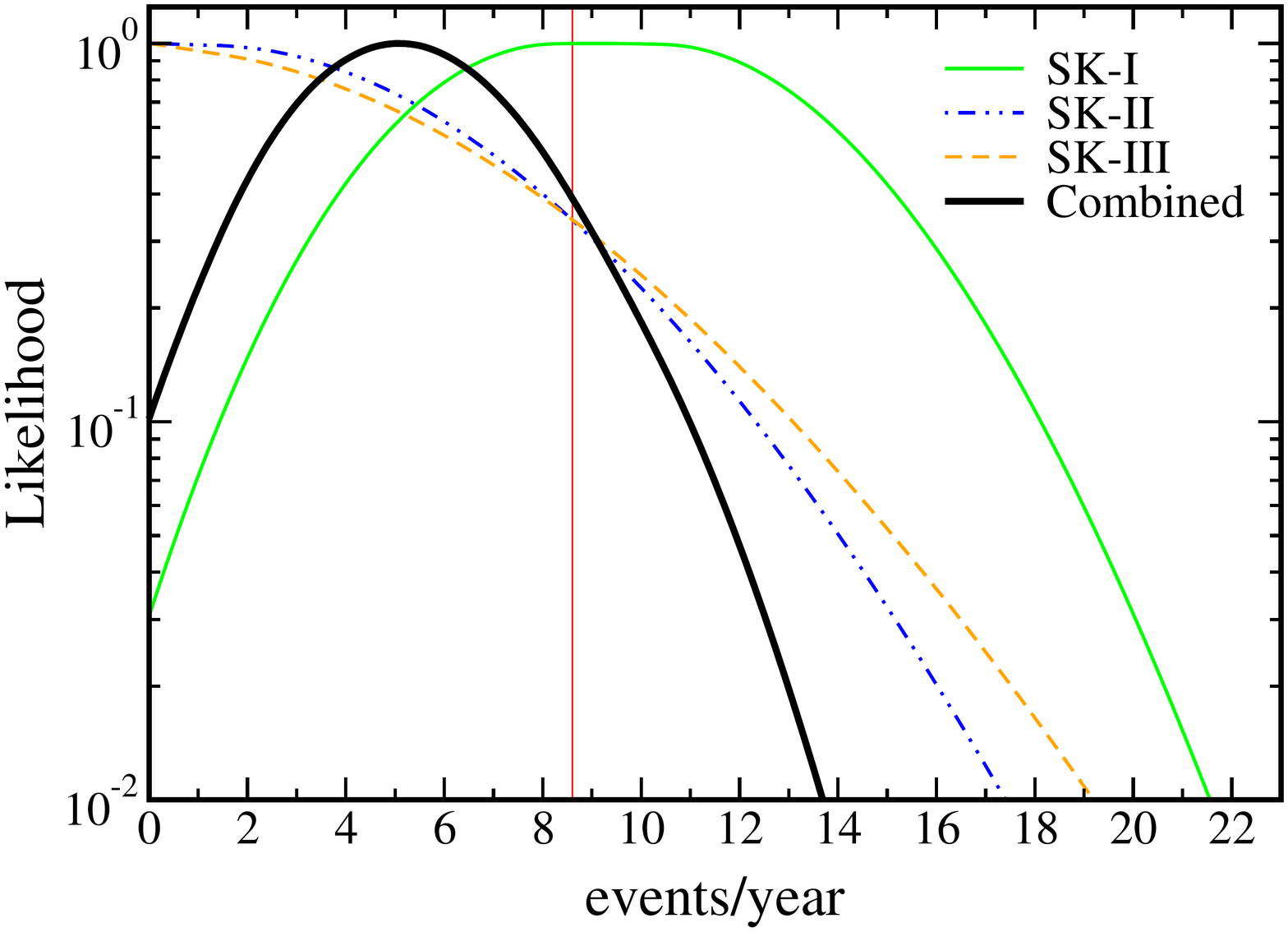} \\[3ex]
\includegraphics[width=0.45\linewidth]{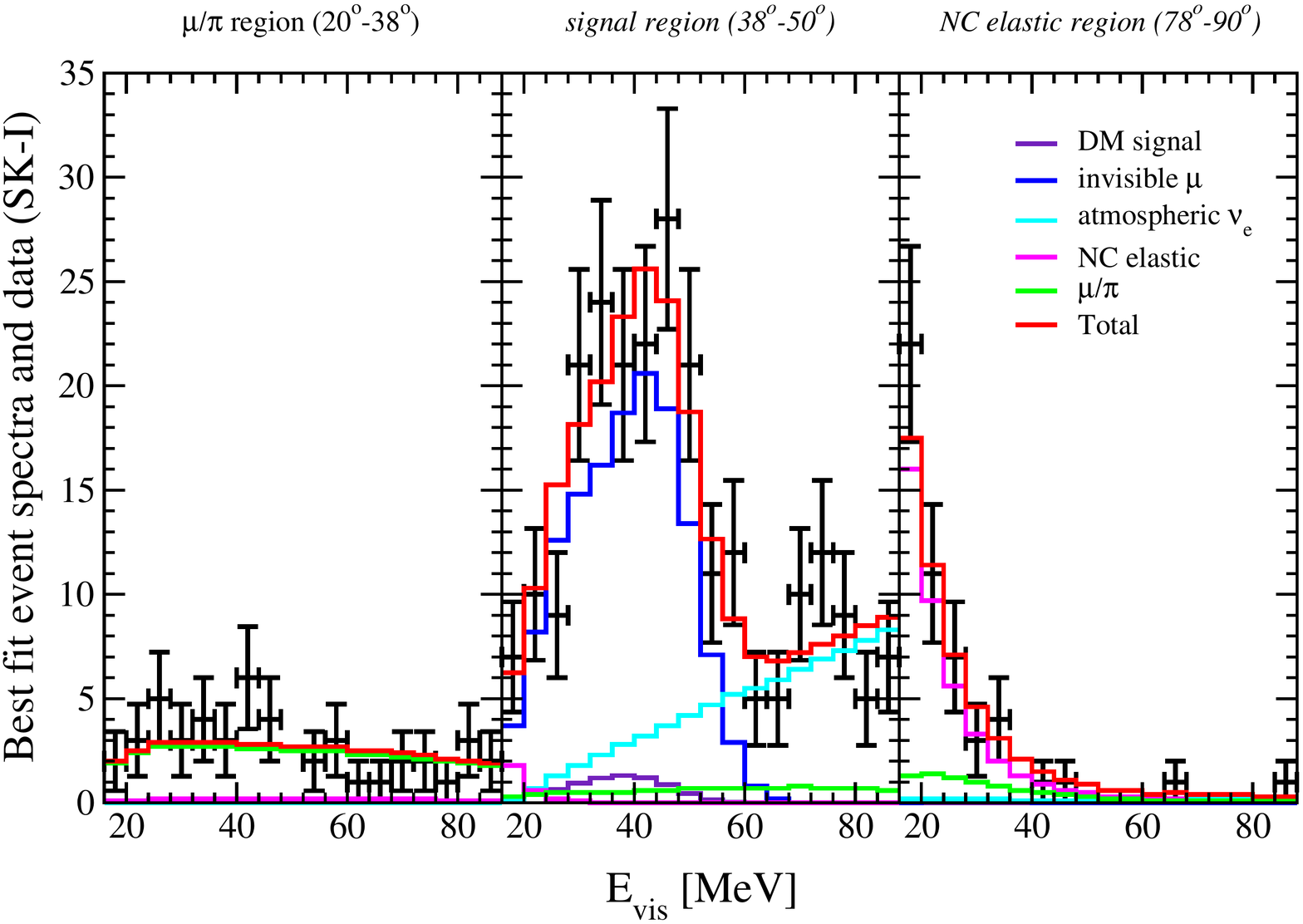}
\hspace{5mm}
\includegraphics[width=0.45\linewidth]{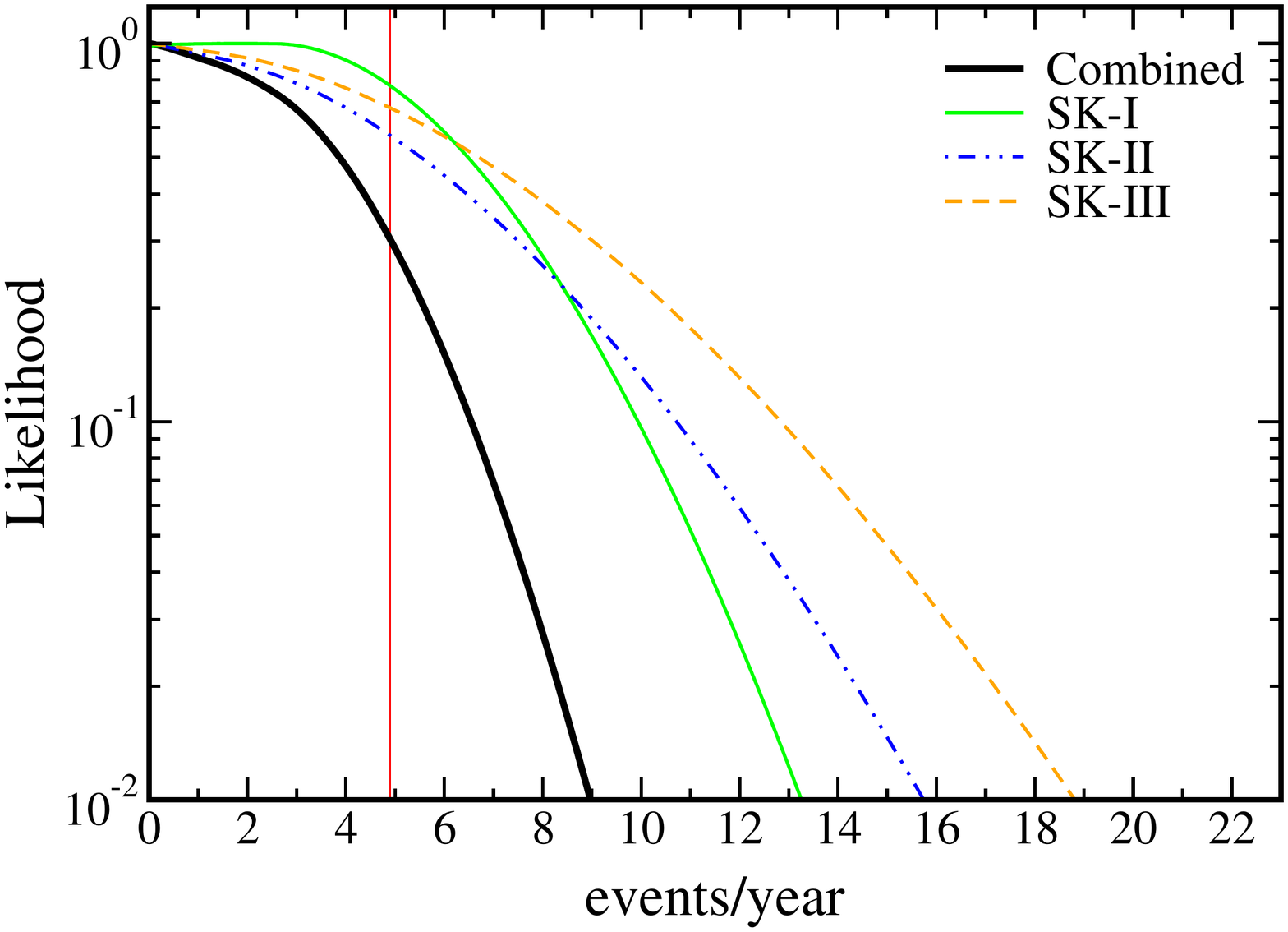}
\caption{\sl \footnotesize \textbf{\textit{Left panels: Best
      fit event spectra and data~\cite{Bays:2011si, Baysthesis}}} for
  SK-I in the three \v{C}erenkov angle regions.  The two main
  backgrounds in the signal region, invisible muons and atmospheric
  $\nu_e$ and $\bar{\nu}_e$, are the blue and cyan histograms,
  respectively.  The $\mu/\pi$ and NC elastic backgrounds are depicted
  by the green and magenta histograms, respectively.  The WIMP signal
  spectra is the violet histogram and the total best fit spectra is
  the red histogram.  \textbf{\textit{Right panels: Likelihood
      normalized to its maximum}} for each SK phase and for the
  combined analysis, as a function of the number of events per year.
  The vertical line represents the 90\%~CL limit for the combined fit. 
  \textbf{\textit{Upper (lower) panels}} are for the case of WIMPs
  annihilations into light quarks ($\mu^+\mu^-$) and $m_\chi =
  6$~GeV.  Note that in the upper panel the combined best fit is 5.1
  signal events.}
\label{fig:fig5}
\end{center}
\end{figure}

We have also included the energy-independent efficiency systematic
error by modifying the likelihood in the way described in
Refs.~\cite{Bays:2011si, Baysthesis}, with a total error different for
each of the data-taking phases.  The final likelihood is maximized for
each SK phase separately, so that it remains as a function of just
$\alpha$, the number of signal events (the best fit event spectra and
data for $m_\chi = 6$~GeV and SK-I are shown in the left panels of
Fig.~\ref{fig:fig5} for the case of WIMPs annihilations into light
quarks and $\mu^+\mu^-$).  Finally, the three likelihoods are
calculated as a function of the number of signal events/year,
$\tilde{\alpha}$, and multiplied (shown in the right panels of
Fig.~\ref{fig:fig5} for the case of WIMPs annihilations into light
quarks and $\mu^+\mu^-$, for $m_\chi = 6$~GeV).  The 90\% confidence
level (CL) limit on the number of signal events/year,
$\tilde{\alpha}_{90}$, is determined by
\begin{equation}
\frac{\int_0^{\tilde{\alpha}_{90}} \mathcal{L}_{\rm{tot}}
  (\tilde{\alpha}) \, d\tilde{\alpha}}{\int_0^{\infty}
  \mathcal{L}_{\rm{tot}} (\tilde{\alpha}) \, d\tilde{\alpha}} = 0.9 ~. 
\end{equation} 
The 90\% CL limit on the scattering cross section, $\sigma_\chi^{90}$,
is then obtained by solving
\begin{equation}
  \Gamma(m_\chi, \sigma_\chi^{90}) \, A^{\rm{tot}} =
  \tilde{\alpha}_{90} \hspace{3mm} , \,  A^{\rm{tot}} \equiv
  \frac{\sum_{SK} A^{SK} t_{SK}}{\sum_{SK} t_{SK}} \, , 
\end{equation}
where $A^{SK}$ is the number of events per WIMPs annihilation for each
SK phase at the detector.

\begin{figure}[t]
\begin{center}
\includegraphics[width=1\linewidth]{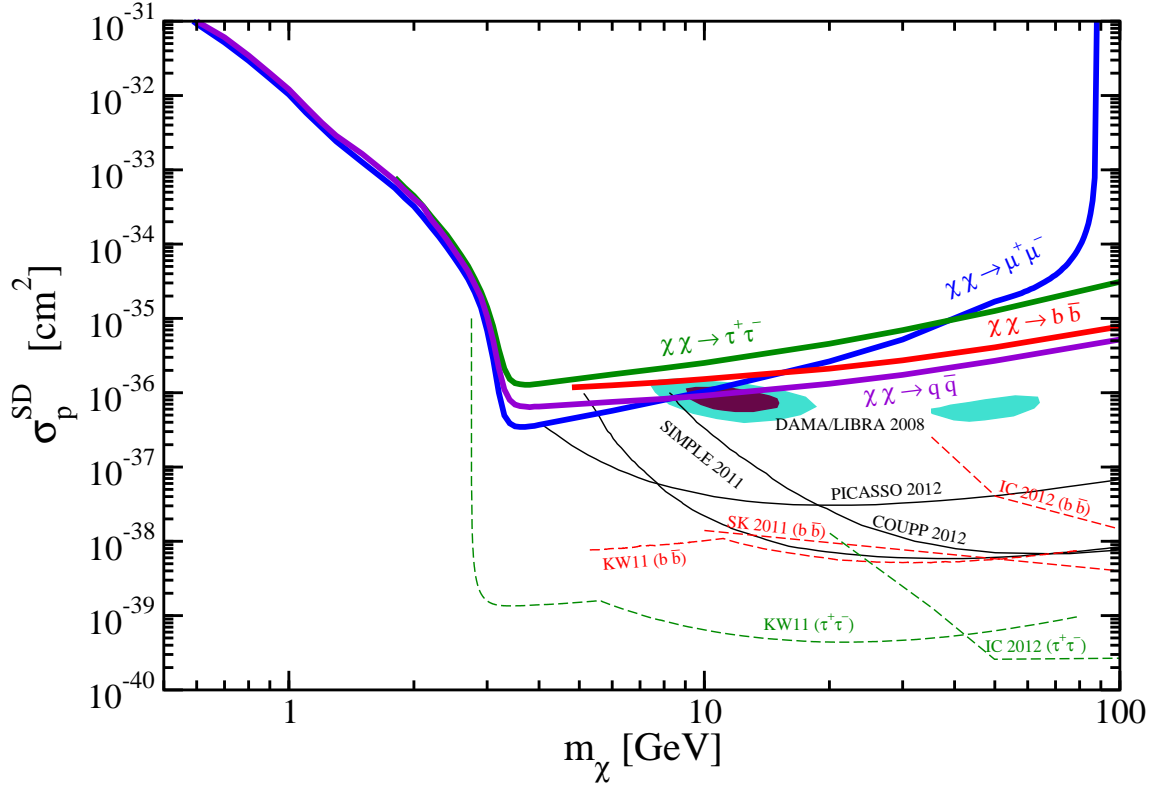}
\caption{\sl \footnotesize \textbf{\textit{Limits on the
      spin-dependent scattering cross section of WIMPs off protons at
      90\%~CL for different annihilation channels}.} The limits from
  SIMPLE~\cite{Felizardo:2011uw}, PICASSO~\cite{Archambault:2012pm}
  and COUPP~\cite{Behnke:2012ys} are shown with black lines.  The
  limits from SK searches of GeV neutrinos (for different data sets)
  are depicted for two annihilation channels~\cite{Tanaka:2011uf,
    Kappl:2011kz}, as well as those from
  IceCube~\cite{Aartsen:2012kia}.  The
  DAMA/LIBRA~\cite{Bernabei:2008yi, Bernabei:2010mq} regions (at
  90\%~CL and 3$\sigma$~CL) are also shown as interpreted in
  Refs.~\cite{Savage:2008er, Savage:2010tg}.
}   
\label{fig:fig6}
\end{center}
\end{figure}

\begin{figure}[t]
\begin{center}
\includegraphics[width=1\linewidth]{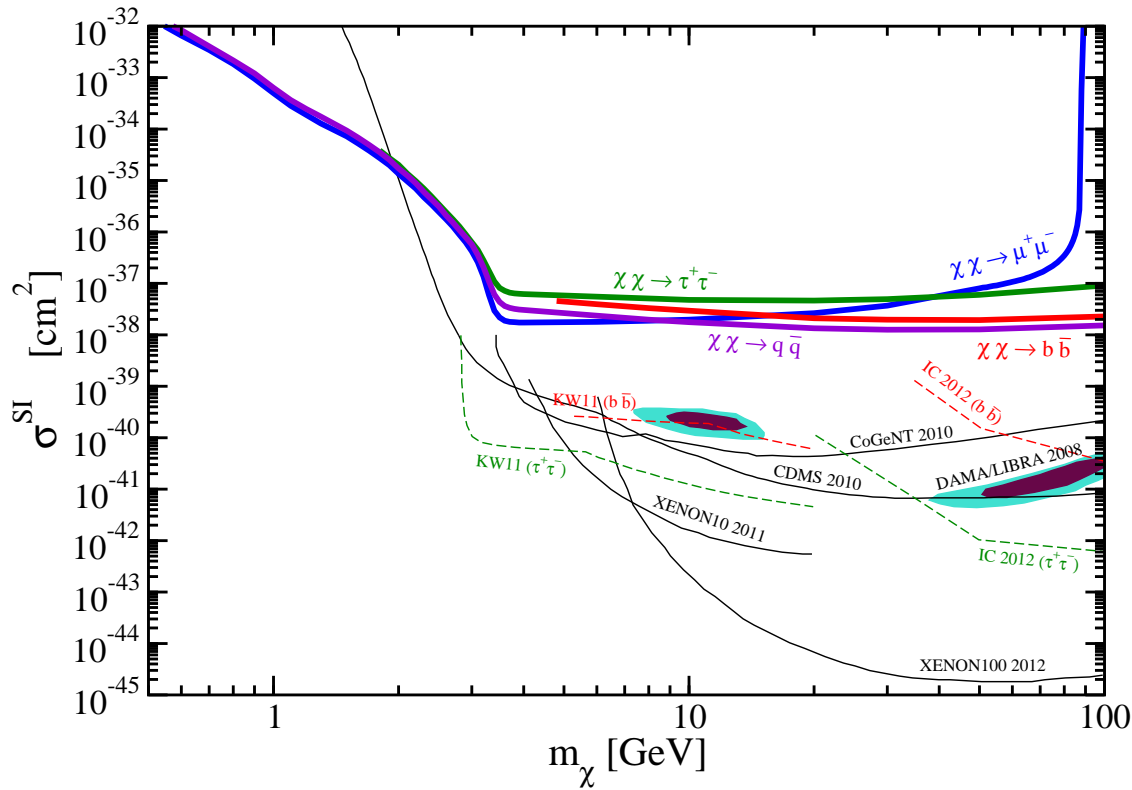}
\caption{\sl \footnotesize \textbf{\textit{Limits on the
      spin-independent scattering cross section of WIMPs at
      90\%~CL for different annihilation channels}.} The limits from
  CoGeNT~\cite{Aalseth:2010vx}, CDMS~\cite{Akerib:2010pv},
  XENON10~\cite{Angle:2011th} and XENON100~\cite{Aprile:2012nq} are
  shown with black lines.  The limits from SK searches of GeV
  neutrinos are depicted for two annihilation
  channels~\cite{Tanaka:2011uf, Kappl:2011kz}, as well as those from
  IceCube~\cite{Aartsen:2012kia}.  The
  DAMA/LIBRA~\cite{Bernabei:2008yi, Bernabei:2010mq} regions (at
  90\%~CL and 3$\sigma$~CL) are also shown as interpreted in
  Refs.~\cite{Savage:2008er, Savage:2010tg}.
}
\label{fig:fig7}
\end{center}
\end{figure}

The results for spin-dependent (off protons) and spin-independent
cross sections are shown in Figs.~\ref{fig:fig6} and~\ref{fig:fig7},
respectively, along with the bounds from direct detection searches and
the analysis of GeV neutrinos from WIMPs annihilations in the Sun
detected at SK.  The most stringent bounds we obtain on the
spin-dependent (spin-independent) cross section are for WIMPs
annihilations into light quarks for $m_\chi>8.4$~GeV
($m_\chi>8.6$~GeV) and for annihilations into $\mu^+\mu^-$ for lower
masses.  Whereas for spin-independent, the limits are a few orders of 
magnitude weaker than the ones obtained with direct searches for
$m_\chi$ above a few GeV, for spin-dependent, they are comparable to
(or even more stringent than) them just below $m_\chi \simeq 10$~GeV.
We note that, in this case, the limits for WIMPs annihilations into
quarks or $\mu^+\mu^-$ exclude part of the DAMA/LIBRA
region~\cite{Bernabei:2008yi, Bernabei:2010mq} at 90\%~CL.  In
addition, it is important to note that below $m_\chi = 4.1$~GeV
($m_\chi = 2.0$~GeV) the limits we obtain for the spin-dependent 
(spin-independent) cross section are more constraining than any direct
detection result and extend to the kinematical limits of each
annihilation channel (not shown on the plots for all the cases), albeit
only reaching relatively large values.  In the optically thick regime
($\sigma_{\rm p}^{\rm SD} \gtrsim 10^{-35}$~cm$^2$ and $\sigma^{\rm
  SI} \gtrsim 3 \times 10^{-37}$~cm$^2$) evaporation is much less effective,
because upscattered WIMPs above the escape velocity have a very short
mean free path and are kept trapped inside the Sun.  Although first
noted in Ref.~\cite{1990ApJ...356..302G}, this is usually overlooked
in the literature and it is usually assumed that WIMPs below a few GeV
cannot get trapped in the Sun.

Although $e^+e^-$ interactions with the solar medium would generate a
modest amount of pions, which subsequently produce MeV neutrinos,
current data are compatible at 90\%~CL with the maximum possible signal
from this channel, obtained for the saturation value of the capture
rate.  Finally, let us note that the limit for the $\mu^+\mu^-$
channel above the evaporation mass follows the dependence of the
capture rate with the WIMP mass, for the initial muons are always
stopped and thus, regardless the WIMP mass, the final number of muons
decaying at rest per WIMPs annihilation is two.  For $m_\chi >
90$~GeV, the capture rate is equal to its geometrical value, so the 
sensitivity decreases drastically (for low masses the effect is more
involved because evaporation is also important).  On the other hand,
for the $\tau^+\tau^-$ channel, below some mass, the number of muons
decaying at rest is smaller because a fraction of taus goes to
electrons and all the produced $\pi^-$ get absorbed without decaying.
However, the passage through the solar medium of the products of tau
decay gives rise to a number of neutrinos that increases (almost
linearly) with energy and hence, with the WIMP mass (a similar
behavior is observed for the case of WIMPs annihilations into quarks).
This explains why these limits cross each other.  Another factor being
the slightly more constraining fit for the $\mu^+\mu^-$ case (see
Fig.~\ref{fig:fig5}).

\section{Conclusions}
\label{sec:Conclusions}
 
The potential signal of GeV neutrinos produced after WIMPs
annihilations in the Sun has been extensively studied so
far~\cite{Kamionkowski:1991nj, Bottino:1991dy, Halzen:1991kh,
  Bergstrom:1996kp, Barger:2001ur, Cirelli:2005gh, Mena:2007ty,
  Blennow:2007tw, Wikstrom:2009kw, Rott:2011fh, Das:2011yr}, although
these searches do not consider annihilations into light quarks,
$\mu^+\mu^-$ or $e^+e^-$, for these channels do not produce GeV
neutrinos.  However, the propagation in the Sun of the products of
WIMPs annihilations into quarks or leptons would always produce pions
and muons that after being stopped would decay at rest.  In this work,
we have considered these MeV neutrinos from pion and muon decay at
rest, which represents a novel way of constraining WIMPs annihilations
in the Sun and, indeed, the only way to set bounds on WIMP
annihilations in the Sun if they are into light quarks (except for the
case of very high WIMP masses), $\mu^+\mu^-$ or $e^+e^-$.  In order to
do so, we have used the SK data on the DSNB~\cite{Bays:2011si,
  Baysthesis} and have performed an analogous analysis.  It is
important to note that taking into account the suppression of the WIMP
evaporation rate from the Sun for large scattering cross sections,
allows us to set bounds for very low WIMP masses, unlike what is
usually assumed within this context.

Our results, Figs.~\ref{fig:fig6} and~\ref{fig:fig7}, show that, mainly
for low WIMP masses and spin-dependent cross sections, these new
limits are competitive with those from direct searches in the few GeV
region and extend to the kinematical limits of each annihilation
channel.  However, SK data do not allow us to set limits on
annihilations into $e^+e^-$ yet.  In addition, note that direct
detection limits are very sensitive to the unknown high-velocity tail
of the WIMPs distribution with uncertainties of up to two orders of
magnitude in the predicted  rates~\cite{Lisanti:2010qx, Mao:2012hf}.
Hence,  in this respect our limits, being little affected by these
systematics, are more robust. 

It is interesting to also point out that, although the angular
distribution of inverse beta-decay events is quite flat, this could in
principle be exploited to further constrain this
signal~\cite{Ueno:2012md}.  In addition, the angular information of
each event could be used to exploit the Earth matter
effect~\cite{Lunardini:2001pb, Peres:2009xe} as well, on an event by
event basis.  Finally,  let us stress that future detectors such as
Hyper-Kamiokande~\cite{Abe:2011ts} could allow us to improve these
limits by up to two orders of magnitude.

\acknowledgments

We thank K.~Bays and C.~Pe\~na-Garay for providing us with useful
information about the SK DSNB analysis and about the SSM,
respectively.  We would also like to thank M.~Vicente-Vacas for
discussions, J.~Casanellas for rising an important point and M.~Sorel
for encouragement.  SPR thanks the Galileo Galilei Institute for 
Theoretical Physics, where parts of this work were carried out, for
hospitality.  NB is supported by the DFG TRR33 `The Dark Universe'.
JMA is supported by the Spanish Grant CONSOLIDER-Ingenio 2010
CSD2008-0037 (CUP) of the MINECO.  SPR is supported by the Spanish
Grant FPA2011-23596 of the MINECO and by the Portuguese FCT through
CERN/FP/123580/2011 and CFTP-FCT UNIT 777, partially funded through
POCTI (FEDER).

\bibliographystyle{utphys}
\bibliography{DMatMeVarXiv2}

\end{document}